\documentclass[fleqn,usenatbib]{mnras}

\usepackage{newtxtext,newtxmath}

\usepackage[T1]{fontenc}

\DeclareRobustCommand{\VAN}[3]{#2}
\let\VANthebibliography\thebibliography
\def\thebibliography{\DeclareRobustCommand{\VAN}[3]{##3}\VANthebibliography}

\usepackage{graphicx}	
\usepackage{amsmath}	
\usepackage{subcaption}
\usepackage{ulem}
\usepackage{natbib}

\title[Kinematic study in the Orion Complex]{Kinematic study of the Orion Complex: analysing the young stellar clusters from big and small structures}

\author[S. Sánchez-Sanjuán et al.]{
Sergio Sánchez-Sanjuán,$^{1}$\thanks{E-mail: sergsanchez@astro.unam.mx}
Jesús Hernández,$^{1}$
Ángeles Pérez-Villegas,$^{1}$
Carlos Román-Zúñiga,$^{1}$
\newauthor
Luis Aguilar,$^{1}$
Javier Ballesteros-Paredes,$^{2}$
Andrea Bonilla-Barroso$^{2}$
\\
$^{1}$Instituto de Astronom\'ia, Universidad Nacional Aut\'onoma de M\'exico, Apartado Postal 106, C. P. 22800, Ensenada, B. C., M\'exico\\
$^{2}$Instituto de Radioastronom\'ia y Astrof\'isica, UNAM, Campus Morelia. PO Box 3-72. 58090. Morelia, Michoac\'an, M\'exico\\
}

\date{Accepted XXX. Received YYY; in original form ZZZ}

\pubyear{2024}

\begin{document}
\label{firstpage}
\pagerange{\pageref{firstpage}--\pageref{lastpage}}
\maketitle

\begin{abstract}
In this work, we analysed young stellar clusters with spatial and kinematic coherence in the Orion star-forming complex. For this study, we selected a sample of pre-main sequence candidates using parallaxes, proper motions and positions on the colour--magnitude diagram. After applying a hierarchical clustering algorithm in the 5D parameter space provided by \textit{Gaia} DR3, we divided the recovered clusters into two regimes: \textit{Big Structures} and \textit{Small Structures}, defined by the number of detected stars per cluster. In the first regime, we found 13 stellar groups distributed along the declination axis in the regions where there is a high density of stars. In the second regime, we recovered 34 clusters classified into two types: 14 as {\it small groups} completely independent from the larger structures, including four candidates of new clusters, and 12 classified as \textit{sub-structures} embedded within 5 larger clusters. Additionally, radial velocity data from APOGEE-2 and GALAH DR3 was included to study the phase space in some regions of the Orion Complex. From the \textit{Big Structure} regime, we found evidence of a general expansion in the Orion OB1 association over a common centre, giving a clue about the dynamical effects the region is undergoing. Likewise, in the \textit{Small Structure} regime, the projected kinematics shows the ballistic expansion in the $\lambda$ Orionis association and the detection of likely events of clusters' close encounters in the OB1 association.

\end{abstract}




\section{Introduction}

Empirical knowledge about star formation shows that most stellar objects tend to form in groups from the same molecular cloud \citep[e.g.,][]{kroupa:1995, lada:2003,Kerr2023}. In their initial evolutionary stages, these stellar populations exhibit a distinct kinematic signature since they have not been disrupted by the Galactic potential. Recently, the study of young stellar clusters (YSCs) has gained significant attention due to the role played by large-scale surveys with unprecedented precision, e.g. the Sloan Digital Sky Survey~\citep[SDSS;][]{almeida:2023} or \textit{Gaia}~\citep{gaia:2016, gaia:2018, gaia:2021}. In this context, the solar neighbourhood is considered an area of interest, where numerous star-forming complexes have been extensively studied~\citep{poppel:1997, bally:2008, zari:2018}.

Particularly, OB associations are gravitationally unbound stellar groups with ages generally younger than 10 Myr~\citep{ambartsumian:1947}. Their study provides a better understanding of the environments and evolutionary stages, which go from stars still embedded inside their natal molecular cloud (typically below $\sim 1$ Myr) to regions more evolved with a significant absence of gas blown out by feedback processes~\citep{briceño:2019, krause:2020}. The bulk of the population in OB associations consists of the low-mass pre-main-sequence (PMS) stars, which leads to a better characterization of the cluster in terms of spatial distribution and kinematics~\citep{briceño:2007, hernandez:2023, wright:2023}. Since these stars preserve the initial movement from the parental molecular cloud, they are good tracers for the star formation history as initially analysed by Hipparcos \citep[e.g.,][]{deZeeuw:1999, hoogerwerf:1999, bouy:2015}. Currently, with the advent of \textit{Gaia}~\citep{gaia:2016, gaia:2018, gaia:2021, gaia:2023},
it is possible to identify a better morphology of the stellar populations within OB associations at different scales using their astrometric information.

The Orion star-forming complex (OSFC) is an outstanding laboratory for assessing YSCs due to its massive composition of gas, active process of star formation and proximity to Earth~\citep{brown:1994, bally:2008,grossschedl:2018, zucker:2020}. During the last decades, many studies have been carried out as more high-quality photometric and spectroscopic data emerge. Some indicate the enrichment of the region in the diversity of star formation processes and stellar properties~\citep{briceño:2019, hernandez:2023, roman-zuñiga:2023}. Similarly, others show the identification of an age distribution along the complex, suggesting a sequential star-forming scenario~\citep{bally:2008, kos:2020}. Additional features, such as the morphological and dynamic characterizations of the stellar and gas components, were also studied~\citep{grossschedl:2018, grossschedl:2021, kounkel:2022_dynamics}.

In terms of clustering, several methods have been used for the identification of groups in star-forming regions from observational data~\citep{román-zuñiga:2008, kos:2020, zari:2019, chen:2020, ratzenbock:2022, ratzenbock:2023, tarricq:2022,  kounkel:2022_perob2}. Particularly, \citet{kounkel:2018_apogee} performed a classification of structures in the OSFC, dividing it into five main groups: Orion A, Orion B, Orion C, Orion D, and $\lambda$ Orionis with a hierarchical clustering algorithm using 6D phase space. A similar approach was carried out by~\citet{roman-zuñiga:2023} to obtain a characterization of the star formation history. However, this approach drastically reduced the star sample due to the lack of radial velocity (RV) information. On the other hand,~\citet{chen:2020} reported a more detailed analysis in the study of sub-structures along the complex, reporting 22 clusters with 2 unsupervised machine learning algorithms using the 5D parameter space from \textit{Gaia} data (three spatial and two proper motion components). However, this work did not apply a selection of PMS candidates to avoid the presence of contamination in the sample. Additionally, all the works mentioned above were performed using \textit{Gaia} DR2~\citep{gaia:2018}, which was superseded by the new release, \textit{Gaia}~DR3.

Moderate to high-resolution spectroscopy provides precise measurements of RV, a parameter required for the 6D phase-space vector, which enables a comprehensive analysis of the star formation history from a dynamic approach. Previous works that combine astrometric and RV information have shown expansion phenomena in the OSFC. Particularly in $\lambda$ Orionis~\citep{kounkel:2018_apogee}, between the Orion C and Orion D structures with a common origin in the geometrical centre of the Barnard's loop~\citep{kounkel:2020_supernova}, and in the surroundings of OBP-Near/Briceño-1 clusters~\citep{swiggum:2021}, which is linked with a void of gas recently studied by~\citet{foley:2023}. The presence of an expansion event can be related either to the occurrence of a supernova or winds, where the gas is swept~\citep{kounkel:2020_supernova} with the additional ingredient that the gravitational potential of the swept molecular could exert an additional pull~\citep{zamora-aviles:2019}. As the evidence of expansion has been identified in two regions of Orion, a deeper analysis is necessary for the whole Orion Complex, searching for likely interactions of the YSCs recovered.

In this paper, we make a general characterization of the sub-structures found in the OSFC using the Hierarchical Density-Based Spatial Clustering of Applications with Noise algorithm~\citep[\textsc{hdbscan};][]{mcinnes:2017, Campello:2015} with the 5D parameter space provided by \textit{Gaia} DR3~\citep{gaia:2021}. Also, adding RV information, we perform a kinematic analysis in phase space to search for group interactions. This work is organized as follows. In section~\ref{sec:data}, we show the data used for our analysis. In section~\ref{sec:methodology}, we present the methodology to identify clusters from the \textit{Gaia}'s observable parameters. In section~\ref{sec:bsr}, we analyse the complex in terms of the \textit{Big Structure} regime and section~\ref{sec:ssr} in terms of the \textit{Small Structure} regime. In section~\ref{sec:merging}, we study likely scenarios of close encounters, followed by conclusions in section~\ref{sec:conclusions}.

\section{Data Sample} \label{sec:data}

In this section, we show the astrometric and spectroscopic data used to identify YSCs in the OSFC and the criteria applied to obtain the star samples.

\subsection{The main sample}
\begin{figure}
    \centering
    \includegraphics[width=0.44\textwidth]{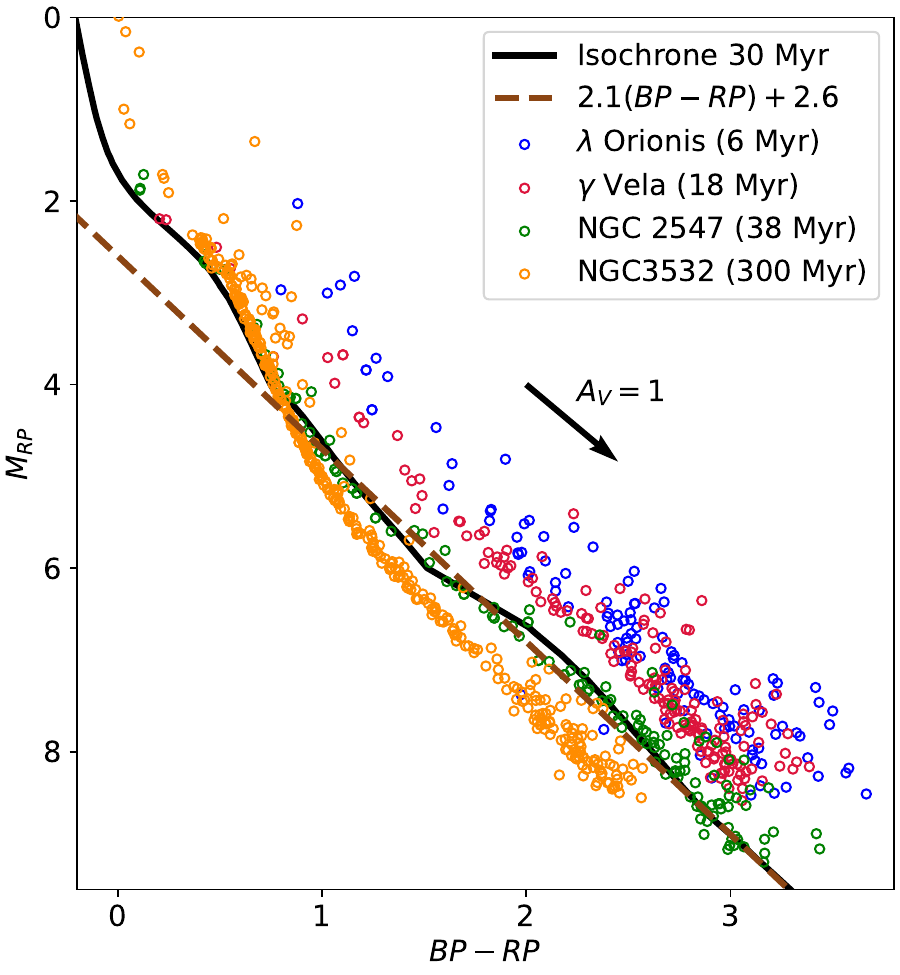}
    \caption{Colour--magnitude diagram with \textit{Gaia} bands including the 30-Myr isochrone from \textsc{parsec} (continuous line) and a linear approximation (dashed line) that matches in the late-type stars regime: $BP-RP >1.0$. Some clusters with different ages were included as comparison: $\lambda$ Orionis (blue), $\gamma$ Vela (red), NGC 2547 (green), and NGC 3532 (orange) as reported in~\citet{jackson:2022}.}
    \label{fig:CMD_cut}
\end{figure}

Initially, young candidates were selected in the Orion region using \textit{Gaia} DR3 data~\citep{gaia:2021, gaia:2023}. The initial catalogue includes sources in the range of $75^{\circ}<\alpha < 90^{\circ}$ and $-14^{\circ}<\delta < 16^{\circ}$. From~\citet{hernandez:2023}, we adopted limits for parallax\footnote{$\varpi$ is corrected by systematics after applying the zero-point bias obtained from the ARI \textit{Gaia} TAP service: https://gaia.ari.uni-heidelberg.de/tap.} ($\varpi$) and total proper motion\footnote{With the module $\mu=(\mu_{\alpha}^{*2}+\mu_{\delta}^{2})^{1/2}.$} ($\mu$) based on the catalogue of spectroscopically confirmed T-Tauri Stars (TTSs) published by~\citet{briceño:2019}. In this work, a $3\sigma$ cut from the median value was used for {$\varpi$} assuming a Gaussian distribution, resulting in a range between 2.0 and 3.6 mas, which corresponds to a distance range between $\sim$277 and $\sim$500 pc. Also, we used a cut of $6\sigma$ in {$\mu$} ($<5.425$~ $\rm{mas}\ yr^{-1}$) to avoid rejecting YSOs with atypical kinematic motions that could be generated by dynamic interactions in the star-forming process \citep[e.g., runaway or walkaway TTSs;][]{hernandez:2023}.

We applied an astrometric quality cut requiring stars with uncertainties of $\sigma_\varpi/\varpi < 0.05$. We also used two astrometric quality parameters: the renormalized unit weight error (RUWE), which evaluates the quality of an astrometric solution, assuming a single-star model for the motion of the source on the sky \citep{lindegren:2021}, and the fidelity parameter defined by \citet{rybizki:2022}, which is related to the quality of the astrometric measurements using machine learning techniques. Based on the recommended limits of these parameters, we apply to the star sample the condition: $RUWE<1.4\ \vee\ fidelity>0.5.$

Finally, we made a colour--magnitude cut for selecting young objects using a linear approximation of the 30-Myr isochrone (dashed line in Figure~\ref{fig:CMD_cut}), defined by the PAdova and TRieste Stellar Evolution Code~\citep[\textsc{parsec};][]{marigo:2017} with \textit{Gaia}'s $R_p$ and $B_p$ photometric bands. By using the isochrone trend for the range $(B_p-R_p)>1.0$, we obtained the following expression:
\begin{equation}
    \label{eq:color-magnitud-cut}
    M_{RP} < 2.6 + 2.1(B_p - R_p).
\end{equation}

\begin{figure*}
    \centering
    \includegraphics[width=0.37\textwidth]{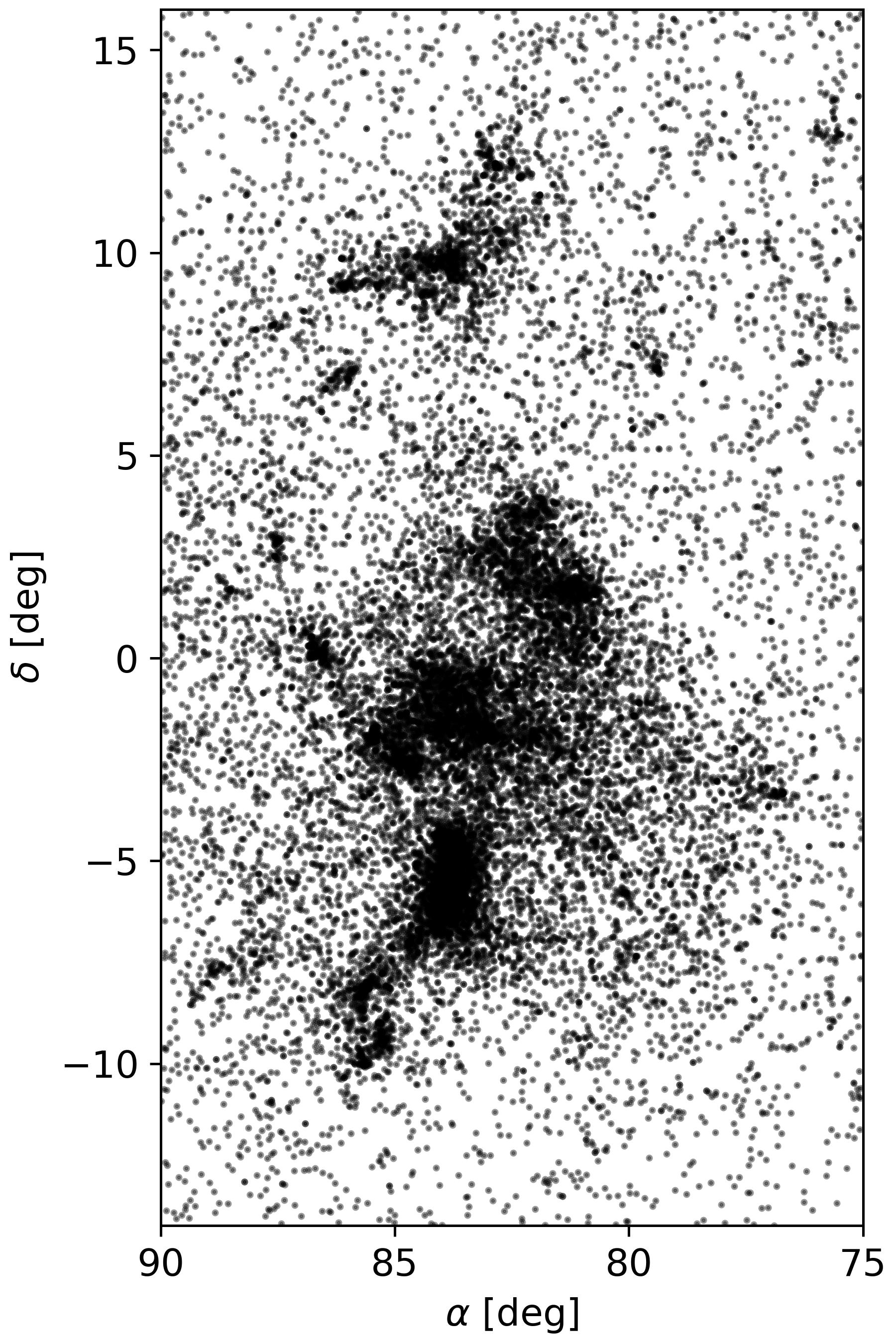}
    \includegraphics[width=0.37\textwidth]{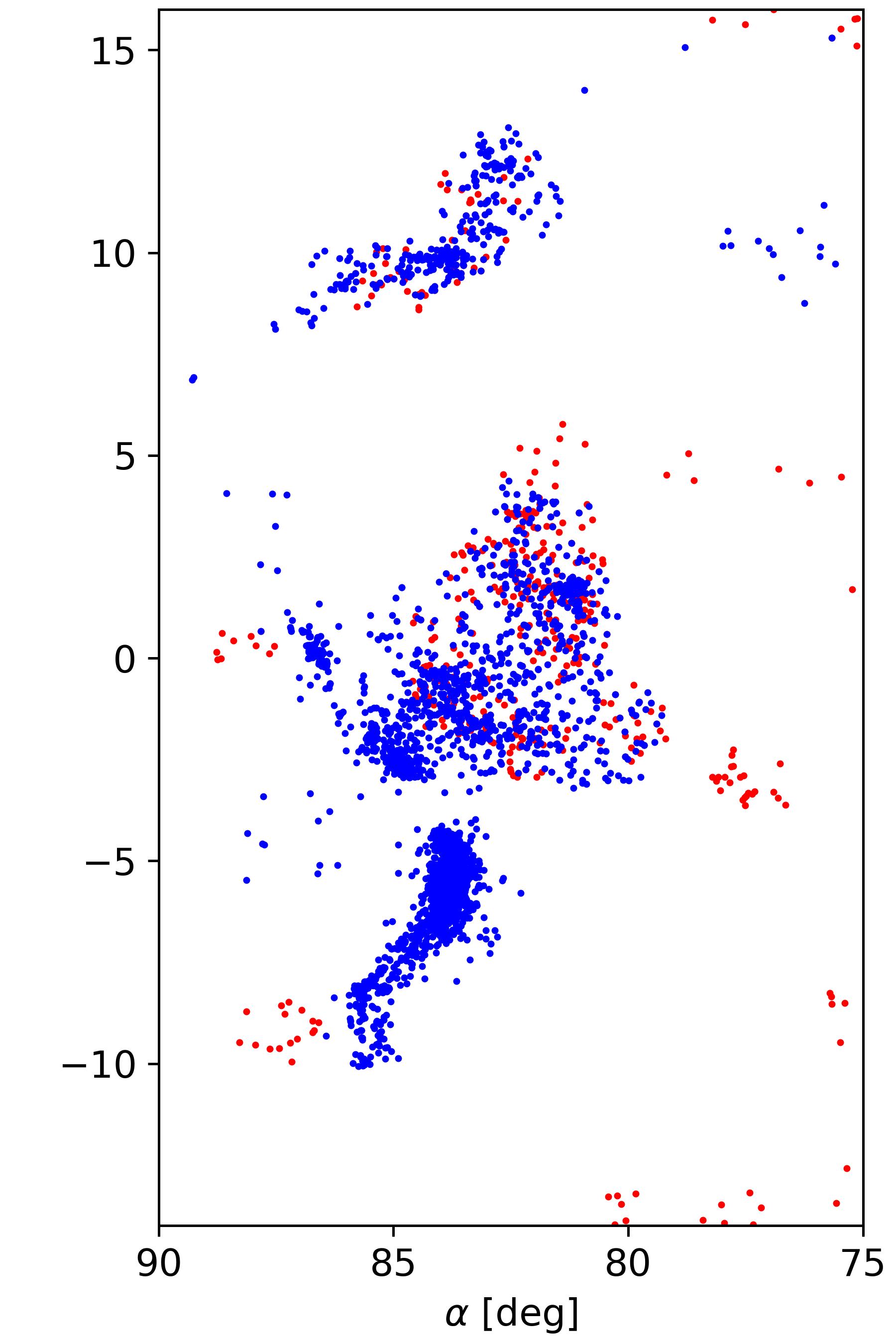}
    \caption{Distribution of the \textit{main sample} (left) and of the \textit{RV sample} (right). In blue we have the stars from APOGEE-2; and in red, the stars in GALAH DR3 but no in APOGEE-2.}
    \label{fig:sample_full}
\end{figure*}

The linear approximation is compared with the position of four YSCs reported by~\citet{jackson:2022}: $\lambda$ Orionis (6 Myr), $\gamma$ Vela (18 Myr), NGC 2547 (38 Myr), and NGC 3532 (300 Myr). Here, we see that equation~\eqref{eq:color-magnitud-cut} matches the isochrone from solar-type to later PMS. For early-type objects, the selection is naturally done by the isochrone itself. Furthermore, since the reddening vector is almost parallel to the empirical linear boundary, this criterion does not miss possible YSO candidates with low or moderate extinctions. After applying all requirements, we ended up with 16,814 that conform the \textit{main sample}. These stars have a spatial distribution in equatorial coordinates, as seen in the left panel of Figure~\ref{fig:sample_full}.

\subsection{Radial velocity sample}

For the RV, high-resolution data from the Apache Point Observatory Galactic Evolution Experiment 2~\citep[APOGEE-2;][]{blanton:2017} and the GALactic Archaeology with the High-efficiency and high-resolution mercator echelle spectrograph~\citep[GALAH-DR3;][]{buder:2021} were compiled. As a first step, each catalogue was cross-matched with the astrometric sample using the \textit{Gaia}'s identifier (\texttt{source\_id}). When both surveys were compared, we found a median positive systematic shift of $\sim$0.2~$\rm{km}\ s^{-1}$ in APOGEE-2 over GALAH-DR3, which was corrected to GALAH-DR3.

The catalogues were filtered considering the reported error ($\sigma_{RV}$) to avoid high-scattered measurements applying the condition: $\sigma_{RV}/{V_{RV}} < 0.2$, and also removing the contamination from outliers by limiting ourselves to RV values between 0 and 50 $\rm{km}\ s^{-1}$, which is a 5$\sigma$ range from the median value of the whole sample. For the final RV sample, we adopted the total sample from APOGEE-2, and for the cases where there is no APOGEE-2 data, we used GALAH-DR3 instead. Therefore, we obtained a total of 3,252 stars with RV information (right panel of Figure~\ref{fig:sample_full}).

\section{Methodology} \label{sec:methodology}

This section explains the methodology used to recover stellar groups from the star sample, executing a clustering algorithm under two regimes: \textit{Big Structures} and \textit{Small Structures}.

\subsection{Clustering algorithm}

To identify YSCs, we used the \textsc{hdbscan} algorithm~\citep{Campello:2015, mcinnes:2017} under the \textit{leaf} method. During its implementation, we used a 5D parameter space composed of the observables given by \textit{Gaia}: ($\alpha$, $\delta$, $\mu_\alpha^*$, $\mu_\delta$, $\varpi$), which is similar to the approach used by~\citet{chen:2020} but with \textit{Gaia} DR2. As the parameters involved combine spatial and kinematical dimensions, we applied a normalization with the median absolute deviation (MAD) from each corresponding dimension to give an equal weight to each parameter. We decided to work in the observable space to avoid the effect of error propagation that would affect the detection of fine structures. Also, we decided to avoid the RV in the parameter space for the clustering algorithm due to the lack of information from the spectroscopic surveys compared to the \textit{main sample}.

\begin{figure*}
    \centering
    \includegraphics[width=0.47\textwidth]{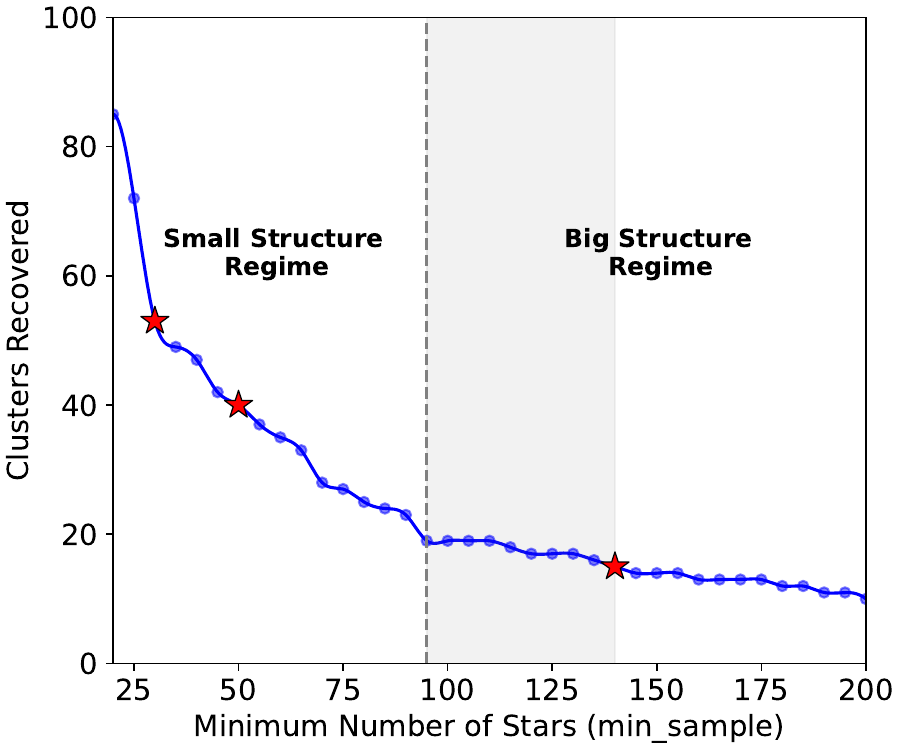}
    \includegraphics[width=0.47\textwidth]{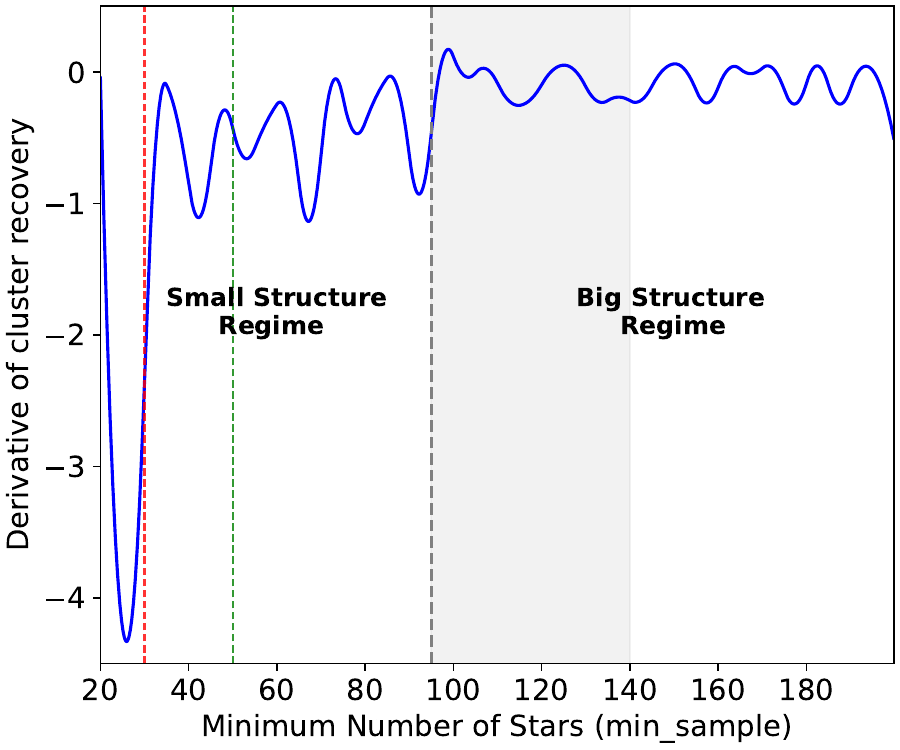}
    \caption{\textit{Left:} number of recovered clusters after running \textsc{hdbscan} for different values of \texttt{min\_samples}. Red stars indicate the cases where \textsc{hdbscan} was used: 140, 50 and 30 stars per cluster. \textit{Right}: the rate of recovered clusters as a function of \texttt{min\_samples}. The red and green lines represent the cases when \texttt{min\_sample} is 30 and 50, respectively, in the \textit{Small Structure} regime. In both plots, the shaded area is a transition region when we move from \textit{Big} to \textit{Small Structures}, and the dashed line is the boundary between both regimes. Finally, a moving average process was applied as a smoothing technique.}
    \label{fig:hdbscan_results}
\end{figure*}

The recovery of YSCs in \textsc{hdbscan} was performed under two regimes: \textit{Big Structures} and \textit{Small Structures}. These regimes were defined based on the number of recovered clusters obtained for \texttt{min\_sample} values between 25 and 200. We illustrate this in the left panel of Figure~\ref{fig:hdbscan_results}. After a visual examination, we observed two different slopes around \texttt{min\_sample} = 95. Above this boundary, clusters are classified into the \textit{Big Structure} regime and below into the \textit{Small Structure} regime. Additionally, we found stellar groups highly spread in the sky projection or $\varpi$ that we defined as extended. The identification of these extended groups is performed after analysing the semi-interquartile range (SIR) over a $5\sigma$ limit in the space distribution of $\alpha$, $\delta$ and $\varpi$ (see appendix~\ref{apendix_b:extended_clusters} for details). The summary of the extended groups is presented in Appendix~\ref{apendix_c:extended}, even though these groups are not included in the upcoming analysis of this work.

For the \textit{Big Structure} regime, we set \texttt{min\_samples} = 140. This choice allows us to identify the five main regions: Orion A, B, C, D, and $\lambda$ Orionis, as outlined by~\citet{kounkel:2018_apogee}, which represent an overall characterization of the whole complex. Values between 140 and 95 define a transition region (grey band in Figure~\ref{fig:hdbscan_results}), where just a few new clusters are recovered, but large groups such as $\lambda$ Orionis start to split into sub-structures.

On the other hand, below \texttt{min\_samples=}95, we see that the number of stellar groups grows faster, as shown in the left panel of Figure~\ref{fig:hdbscan_results}, indicating the presence of smaller structures and the division of larger groups into minor associations. Also, the right panel of Figure~\ref{fig:hdbscan_results} shows that the derivative of cluster recovery has more fluctuations for \texttt{min\_samples} smaller than 95, which states that the number of groups increases with more variable rates. Thus, for the \textit{Small Structure} regime, we explored \texttt{min\_samples} < 95 and decided to set the parameter at 50 to recover most of the clusters reported in the literature. Below this limit, the number of clusters increases dramatically, with most of them identified as extended groups. Nevertheless, to obtain the small known clusters B30 and B35 that belong to $\lambda$ Orionis, we must apply the clustering algorithm further down the limit of 50 stars per cluster. Therefore, to avoid the effects of the extended groups, we analysed $\lambda$ Orionis separately from the rest of the main sample using the parameter \texttt{min\_samples=}30 as a new threshold.

In both regimes, we chose the stars with membership probability > 0.5 as members. After every result, a visual inspection was performed to identify the real groups alongside their spatial distribution compared with the literature.

\subsection{Velocity and position space}

The transformation from the observable space to the Cartesian phase space was performed using the \textsc{astropy} library ~\citep{astropy:2022}. We determined the position vector ($X_h, Y_h, Z_h$) for each star in the Heliocentric-Cartesian reference frame with the $\bmath{Z}$ component pointing toward the North Galactic Pole, the $\bmath{X}$ component pointing toward the Galactic Centre, and the $\bmath{Y}$ component pointing in the direction of Galaxy rotation. For distance determination, we used the inverse of parallax ($1/\varpi$) since the selected targets have a fractional error below 5\% ~\citep{Bailer2015, Bailer2021}. From proper motions and distance in the \textit{main sample} with the inclusion of the \textit{RV sample}, we calculated the heliocentric velocity vector $\vec{v}_h=(U_h,V_h,W_h)$, applying the correction for the Sun's peculiar velocity with the values $\vec{v}_\odot=(U_\odot, V_\odot, W_\odot)=(11.1, 12.24, 7.25)$ $\rm{km}\ s^{-1}$ defined by~\citet{schonrich:2010} to convert with respect to the local standard of rest (LSR).

We used the Monte Carlo error propagation method to determine the positions and velocities for each star, assuming a Gaussian distribution for the six observable parameters with a sampling of 10,000 values. Subsequently, we calculated each cluster's centroid and velocity vector by taking the median value and using the median absolute deviation as the measure of the uncertainty from the sampled stellar data set. The Cartesian phase space will then be used to analyse the kinematic projection of detected clusters in the \textit{Big Structure} regime (see section~\ref{sec:bsr}) and the search for likely future close encounters (see section~\ref{sec:merging}) in the~\textit{Small Structure} regime.

\section{Big Structure Regime} \label{sec:bsr}

\begin{figure*}
    \centering
    \includegraphics[width=0.95\textwidth]{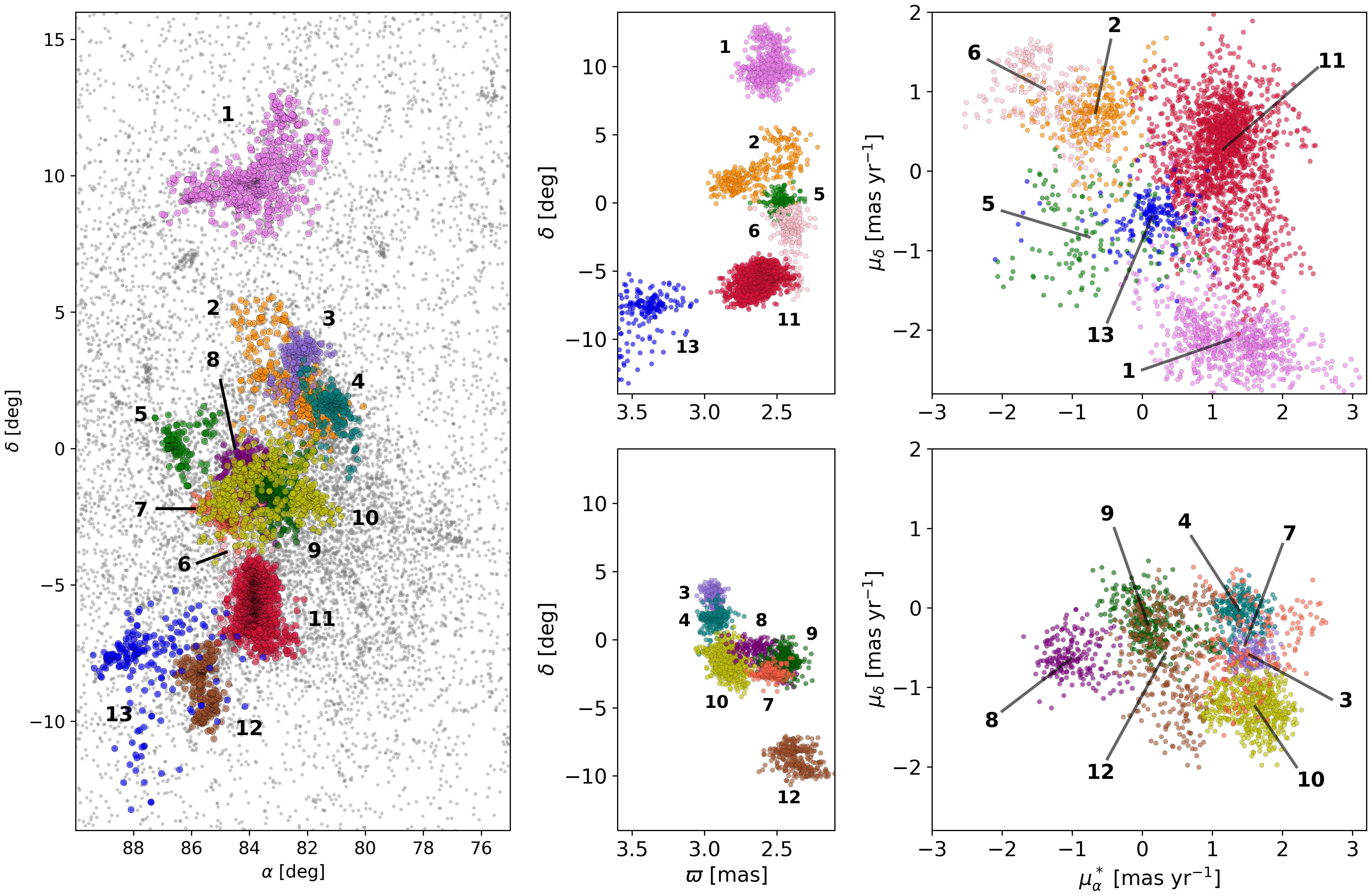}
    \caption{Distribution of \textit{Big Structures} recovered in the Orion Complex. \textit{Left panel}: position of the structures in the sky plane compared with the \textit{main sample} (grey). \textit{Middle panels}: distribution in parallax versus declination. \textit{Right panels}: vector--point diagram. Note that in the central and right panels, the clusters were divided arbitrarily into two groups for better visualization. The numbers are the identifiers as seen in Table~\ref{table:BS}.}
    \label{fig:clusters_BS}
\end{figure*}

{\renewcommand{\arraystretch}{1.24}
\begin{table*}
\centering
\caption{Median observational parameters of the \textit{Big Structures} recovered in the OSFC with the 16th and 84th percentiles. The cluster number is used for identification in Figure~\ref{fig:clusters_BS}. $N_T$ corresponds to the total number of stars from the \textit{main sample} and with radial velocity in parenthesis. The last column shows the references that have studied the clusters using \textit{Gaia} data.}
\label{table:BS}
\begin{tabular}{clcccccccl}
\hline
Cluster & \multicolumn{1}{c}{Name} & \begin{tabular}[c]{@{}c@{}}$\bar{\alpha}$\\ {(}deg{)}\end{tabular} & \begin{tabular}[c]{@{}c@{}}$\bar{\delta}$\\ {(}deg{)}\end{tabular} & \begin{tabular}[c]{@{}c@{}}$\bar{\varpi}$\\ {(}mas{)}\end{tabular} & \begin{tabular}[c]{@{}c@{}}$\bar{\mu}_\alpha^*$\\ {(}$\rm{mas}\ yr^{-1}${)}\end{tabular} & \begin{tabular}[c]{@{}c@{}}$\bar{\mu}_\delta$\\ {(}$\rm{mas}\ yr^{-1}${)}\end{tabular} & \begin{tabular}[c]{@{}c@{}}$\overline{RV}$\\ {(}$\rm{km}\ s^{-1}${)}\end{tabular} & \begin{tabular}[c]{@{}c@{}}$N_T$\\ (with RV)\end{tabular} & References \\ \hline\hline
1  & $\lambda$ Ori   & 83.75$^{+1.19}_{-1.01}$ & 9.83$^{+1.26}_{-0.72}$ & 2.55$^{+0.08}_{-0.07}$ & 1.25$^{+0.59}_{-0.59}$ & $-$2.12$^{+0.40}_{-0.33}$ & 27.26$^{+2.64}_{-1.63}$ & 773 (250) & 1, 2, 3  \\
2  & Ori-North$^*$       & 82.39$^{+1.21}_{-0.71}$ & 2.09$^{+1.89}_{-1.05}$ & 2.62$^{+0.17}_{-0.24}$ & $-$0.67$^{+0.33}_{-0.34}$ & 0.73$^{+0.27}_{-0.29}$ & 30.63$^{+2.24}_{-1.70}$ & 305 (59) & 2, 3, 4 \\
3  & Briceño-1A$^{\dagger}$      & 82.34$^{+0.28}_{-0.41}$ & 3.43$^{+0.36}_{-0.92}$ & 2.94$^{+0.03}_{-0.04}$ & 1.53$^{+0.18}_{-0.20}$  & $-$0.59$^{+0.17}_{-0.15}$ & 20.35$^{+2.07}_{-2.67}$ & 175 (36) & 2, 3 \\
4  & Briceño-1B$^{\dagger}$      & 81.13$^{+0.44}_{-0.34}$ & 1.62$^{+0.29}_{-0.61}$ & 2.93$^{+0.06}_{-0.05}$ & 1.38$^{+0.23}_{-0.17}$  & $-$0.03$^{+0.15}_{-0.19}$ & 20.83$^{+1.22}_{-1.40}$ & 237 (55) & 2, 3 \\
5  & Ori-East        & 86.54$^{+0.25}_{-0.50}$ & 0.13$^{+0.49}_{-0.44}$ & 2.47$^{+0.05}_{-0.05}$ & $-$0.72$^{+0.94}_{-0.58}$ & $-$0.83$^{+0.53}_{-0.44}$ & 28.11$^{+1.93}_{-1.42}$ & 144 (65) & 1, 2 \\
6  & OBP-Far         & 83.96$^{+0.56}_{-0.45}$ & $-$1.85$^{+0.98}_{-2.89}$ & 2.41$^{+0.06}_{-0.06}$  & $-$1.40$^{+0.51}_{-0.41}$ & 1.03$^{+0.36}_{-0.47}$ & 30.52$^{+3.05}_{-1.82}$ & 223 (31) & 2   \\
7  & $\sigma$ Ori    & 84.78$^{+0.31}_{-0.23}$  & $-$2.55$^{+0.37}_{-0.19}$ & 2.54$^{+0.06}_{-0.06}$  & 1.46$^{+0.64}_{-0.39}$ & $-$0.47$^{+0.53}_{-0.43}$ & 30.84$^{+1.45}_{-1.98}$ & 207 (110) & 1, 2 \\
8  & OBP-b           & 84.00$^{+0.45}_{-0.54}$  & $-$0.85$^{+0.53}_{-0.90}$ & 2.61$^{+0.11}_{-0.14}$  & $-$1.02$^{+0.35}_{-0.27}$ & $-$0.64$^{+0.27}_{-0.20}$ & 31.12$^{+4.04}_{-5.40}$ & 196 (31)  & 2, 3  \\
9  & OBP-d           & 83.15$^{+0.32}_{-0.41}$  & $-$1.65$^{+0.63}_{-0.40}$ & 2.45$^{+0.07}_{-0.05}$  & 0.08$^{+0.39}_{-0.24}$  & $-$0.21$^{+0.31}_{-0.26}$ & 30.84$^{+1.04}_{-1.00}$ & 301 (58) & 2, 3 \\
10 & OBP-Near        & 83.55$^{+1.20}_{-1.53}$  & $-$1.77$^{+1.07}_{-0.69}$ & 2.83$^{+0.08}_{-0.07}$  & 1.61$^{+0.28}_{-0.43}$  & $-$1.24$^{+0.24}_{-0.23}$  & 22.77$^{+2.06}_{-2.57}$ & 534 (133) & 2, 3 \\
11  & ONC            & 83.82$^{+0.25}_{-0.30}$  & $-$5.58$^{+0.75}_{-0.84}$ & 2.62$^{+0.08}_{-0.08}$  & 1.16$^{+0.38}_{-0.44}$  & 0.28$^{+0.49}_{-0.74}$  & 26.98$^{+3.53}_{-2.55}$ & 1816 (828) & 1, 2, 3 \\
12  & Ori-South$^{\ddagger}$      & 85.62$^{+0.27}_{-0.33}$  & $-$8.40$^{+0.48}_{-1.19}$ & 2.38$^{+0.07}_{-0.11}$  & 0.33$^{+0.38}_{-0.37}$  & $-$0.57$^{+0.54}_{-0.70}$ & 21.44$^{+1.80}_{-1.95}$ & 331 (117) & 2, 3  \\
13  & Orion Y        & 87.84$^{+0.70}_{-1.44}$  & $-$7.59$^{+0.70}_{-1.82}$ & 3.38$^{+0.14}_{-0.09}$  & 0.14$^{+0.32}_{-0.32}$  & $-$0.57$^{+0.24}_{-0.24}$ & 16.38$^{+2.42}_{-0.11}$ & 189 (3) & 1, 2, 3  \\ \hline
\multicolumn{10}{l}{\footnotesize{$^*$ Ori-North is composed by two substructures: $\omega$-Ori and ASCC 20 as seen in section~\ref{sec:sub_structures}.}}\\
\multicolumn{10}{l}{\footnotesize{$^\dagger$ Briceño-1A and Briceño-1B were identified as a unique structure by~\citet{chen:2020} under the name Briceño-1.}}\\
\multicolumn{10}{l}{$^\ddagger$ Ori-South is composed by two sub-structures: L1641S and L1647 as seen in section~\ref{sec:sub_structures}.}\\
\multicolumn{10}{l}{\textbf{References:} (1)~\citet{kounkel:2018_apogee}, (2)~\citet{chen:2020}, (3)~\citet{swiggum:2021}, (4)~\citet{kos:2020}.}\\
\end{tabular}
\end{table*}
}

\begin{figure*}
    \centering
    \includegraphics[width=0.98\textwidth]{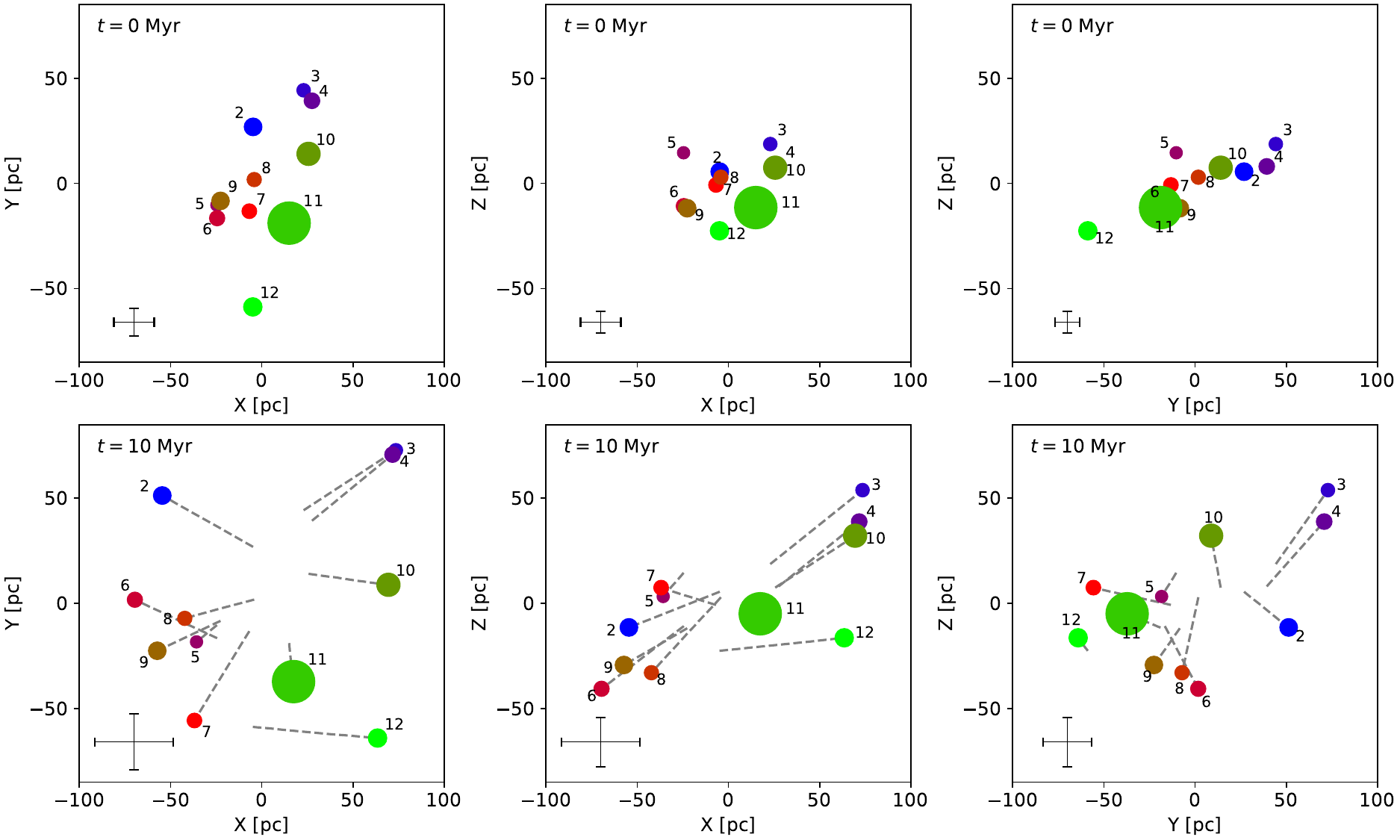}
    \caption{Kinematical behaviour of the entire complex in the XY, XZ, and YZ projections after subtracting the median value of the positions and velocities to the structures, excluding the clusters 1 and 13. The top panels correspond to the current position while bottom panels show the position of the clusters after 10 Myr. The size is related to the number of stars ($N_T$) contained in each cluster and the coordinate (0,0) is the mean of cluster centroids of the whole OB1 association. In the lower left corner of each panel is shown the mean uncertainty for the initial and final position.}
    \label{fig:expansion}
\end{figure*}

For this regime, we identified the regions in the OSFC with a high number of stars sharing the same spatial and kinematic distributions.
To do that, we used the \textsc{hdbscan} algorithm looking for groups of at least 140 stars. As a result, we found 13 stellar groups. Table~\ref{table:BS} 
provides the median value for the observational parameters and the number of identified members for each group. The algorithm also recovered two additional clusters classified as extended, which were discarded since they are highly scattered in their spatial components (see Appendix~\ref{apendix_b:extended_clusters}).

\subsection{Clusters recovered}

In Figure~\ref{fig:clusters_BS}, we show the distributions of the \textit{Big Structures} in terms of the \textit{Gaia} 5D parameter space: $\alpha$ versus $\delta$, $\delta$ versus $\varpi$, and the vector--point diagram. Spatially, we find well-defined clusters distributed along the complex with several superimposed groups where the Orion OB1 association is located, but successfully discriminated in terms of $\varpi$. We also see spatially isolated groups in the northern and southern sectors. Taking the vector--point diagram in Figure~\ref{fig:clusters_BS}, all groups distribute inside a significant range of proper motion values that can be enclosed in a box between $-2.5<\mu_\alpha^*<3$ and $-3<\mu_\delta<2$ $\rm{mas}\ s^{-1}$, with the majority of stars having positive values of $\mu_\alpha^*$.

In the northern sector, we find $\lambda$ Orionis (cluster 1) as a relatively isolated group (the second most populous group). It shows a distance of $399\pm 11$ pc, which closely agrees with the value previously calculated by~\citet{kounkel:2018_apogee} and~\citet{chen:2020}. When the median value of proper motions is subtracted from the whole set of stars contained in the group, we see the effect of ballistic expansion noted by~\citet{kounkel:2018_apogee} as evidence of a possible supernova event $\sim$6 Myr ago. Alternatively, \citet{zamora-aviles:2019} suggest that this kinematic behaviour can also be produced by a gravitational feedback mechanism in which ionizing radiation from massive stars can expel gas from the centre, making the gravitational potential shallower due to inside-out mass removal.

In the Orion OB1a subassociation, three groups share spatially the same sky region: Ori-North, Briceño-1A and Briceño-1B (clusters 2 to 4), with the last two receiving their names from a unique structure identified by~\citet{chen:2020} based on the studies from~\citet{briceño:2007}. Ori-North is located at 381$\pm$33 pc, which is around $\sim$40 pc further than Briceños' clusters with similar distances around 340$\pm$5 and 341$\pm$6 pc, respectively. However, Ori-North has an extended distribution of sources in the line of sight that can be witnessed in the $\varpi$ component, showing a distance dispersion of $\pm 32$ pc, the highest dispersion of all large structures identified. One important result is the recovery of both Briceño-1 clusters as separate groups in \textsc{hdbscan}. Previous works such as~\citet{chen:2020} and ~\citet{swiggum:2021}, with different algorithms, reported only one structure with \textit{Gaia} DR2, indicating the sensibility of our method and the precision in the DR3 catalogue. Additionally, in terms of proper motions, both clusters exhibit comparable kinematics, particularly in the $\mu_{\alpha*}$ component, suggesting a similar star formation history.

In the Orion Belt Population (OBP), also referred as the Orion OB1b subassociation, we identified the most populous area within the complex, where clusters associated with the Orion C and Orion D regions \citep[as defined by][]{kounkel:2018_apogee} share similar locations in the sky. In total, five clusters were recovered, spanning a $\varpi$ range between 2.3 and 3.0 mas. The closest one is the OBP-near group (cluster 10), located at $359\pm 10$ pc and first reported by~\citet{chen:2020}. Behind OBP-near, we recovered the four remaining groups: OBP-b, OBP-d, $\sigma$ Orionis, and OBP-far (clusters 6 to 9)~\citep{chen:2020, swiggum:2021} with calculated distances of $401\pm 9$, $389\pm 18$, $416\pm 10$ and $422\pm 11$ pc respectively.

On the East side of the Orion Belt, we found the Ori-East group (cluster 5) with a median distance of $411\pm 8$ pc, which is assigned to the Orion B structure in~\citet{kounkel:2018_apogee}. \citet{chen:2020} did not recover this group in one of the clustering algorithms they implemented. For this reason, they decided to avoid it in their final list. If we take the average proper motion vector, we identify an apparent movement toward the Orion Belt, which agrees with the general behaviour reported for Orion B~\citep{kounkel:2018_apogee}.

In the southern region of the complex, we identified three clusters with names: ONC, Ori-South, and Orion Y (clusters 11 to 13). The first one is associated with the Orion Nebula Cloud~\citep{alves:2012}, which shows the most populous cluster in this work with 1,816 stars at a distance of $387\pm 12$ pc. This cluster shows a highly spread proper motion distribution, more extensive in $\mu_\delta$ with a range between $-2.0$ and 2 $\rm{mas}\ yr^{-1}$. Ori-South is the combination of L1641S and L1647 groups; both reported by~\citep{chen:2020}. Finally, Orion Y is the closest group identified in the \textit{Big Structure} regime with a distance of $298\pm 9$ pc.

\subsection{Kinematical analysis}

\begin{figure}
    \centering
    \includegraphics[width=0.47\textwidth]{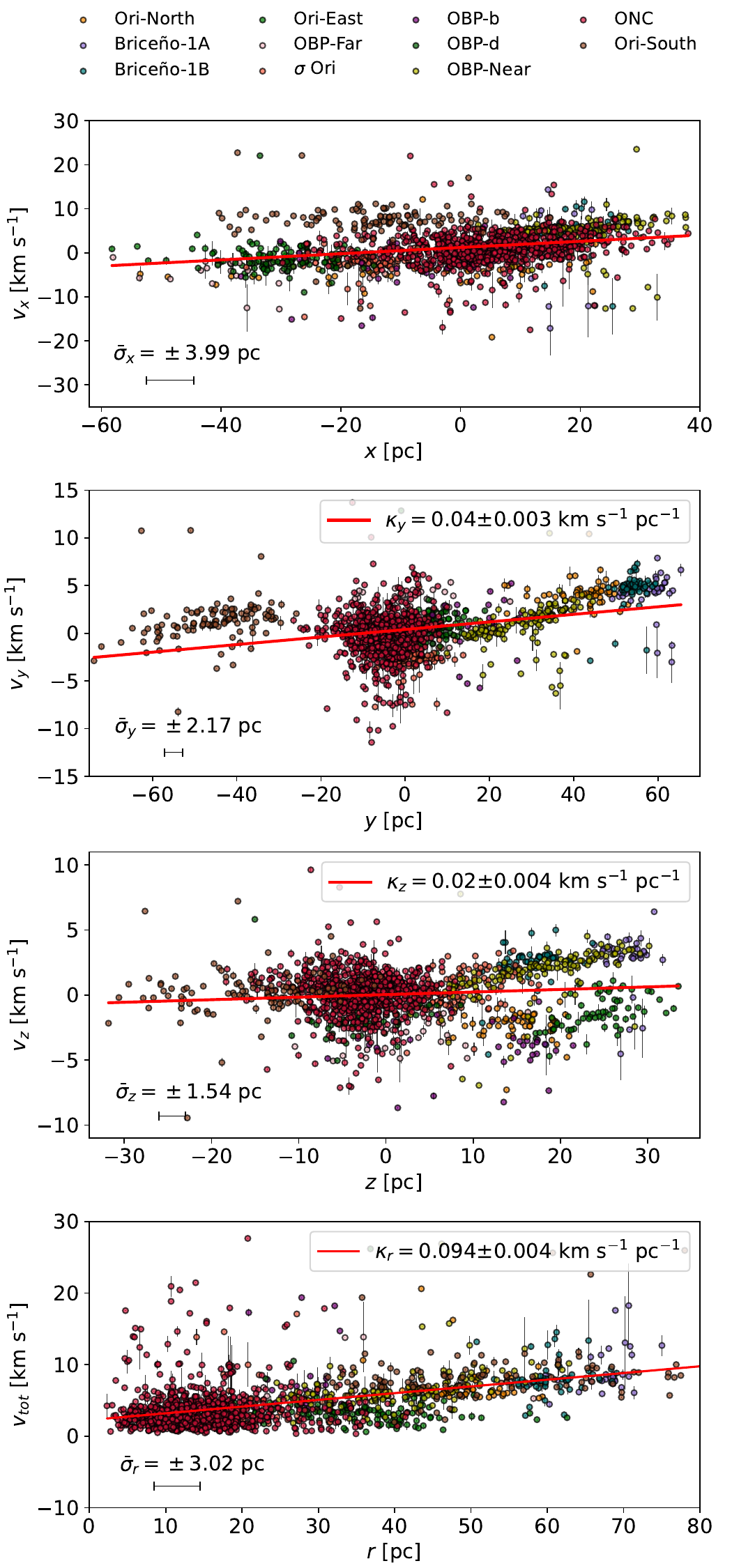}
    \caption{Position-velocity profiles from the reference frame of the OB1 association. We included the regression lines estimated for each panel with their corresponding error. Average error in the space dimension is shown on the lower-right side with the mean value.}
    \label{fig:Profile_Pos_Vel}
\end{figure}

\begin{figure*}
    \centering
    \includegraphics[width=0.88\textwidth]{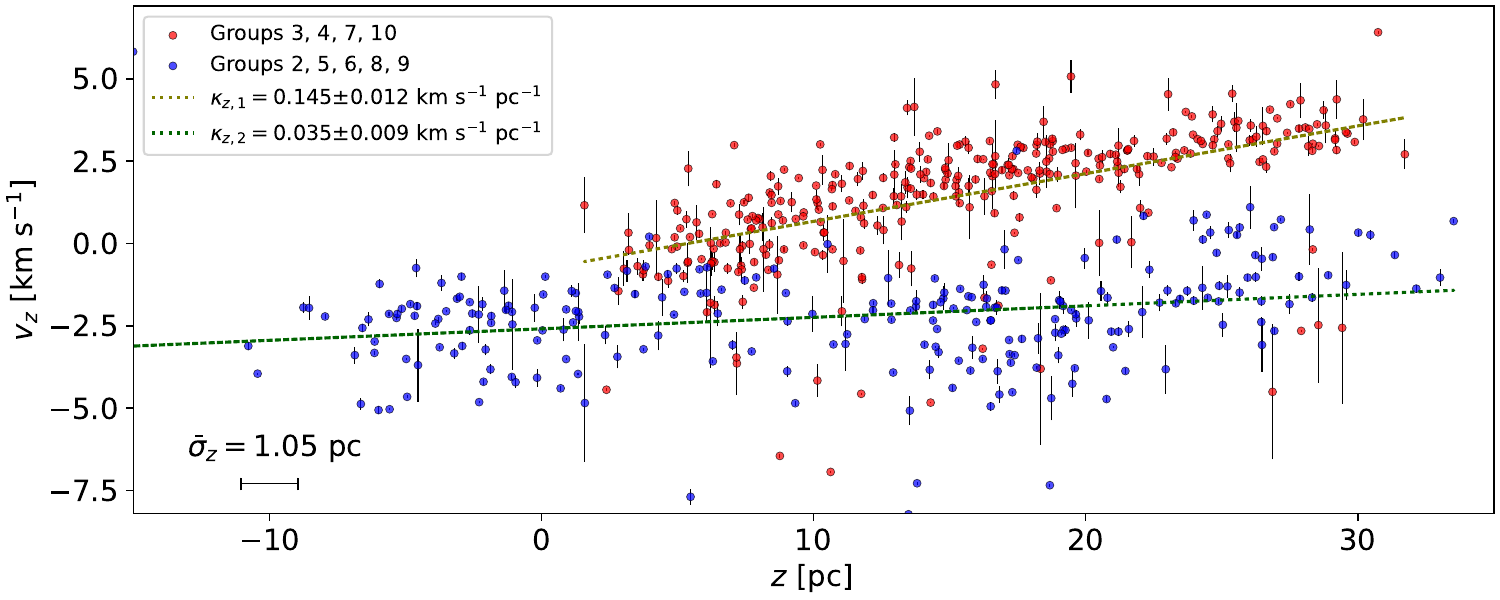}
    \caption{Position-velocity profile for the OB1 association in the Z--W projection. The bifurcation in the $v_z$ axis corresponds to a double linear trend caused by the movement of two sets of clusters: 3, 4, 7 and 10 in $+v_z$ and 2, 5, 6, 8 and 9 in $-v_z$. The cluster number is related to the information shown in Table~\ref{table:BS}}
    \label{fig:zw_bifurcation}
\end{figure*}

Taking the centroid and the median of the velocity vector of each group with respect to the LSR, we extrapolate the trajectories that clusters would trace as a function of time to see the likely future positions from their current locations. For consistency, we removed $\lambda$ Orionis and Orion Y since they are isolated groups, leaving this analysis to the Orion OB1 association. In Figure~\ref{fig:expansion}, we show the kinematic projections of the whole system for the next 10 Myr, assuming rectilinear motion and ignoring gravitational interactions for simplicity. Here, we have chosen the centre coordinate as the mean of the calculated cluster centroids recovered in the OB1 association.

At first glance, we identify a general expansion of the complex in most groups with two remarkable exceptions: OBP-Far and Ori South. Evaluating the tendency of the data in position-velocity profiles, we observed a correlation in all the projections, allowing us to fit a linear model, as depicted in Figure~\ref{fig:Profile_Pos_Vel}. The values of the slopes in each case are $\kappa_x=0.070\pm 0.006$, $\kappa_y = 0.040\pm 0.003$ and $\kappa_z = 0.020\pm 0.004$ $\rm{km\ s^{-1}\ pc^{-1}}$. From these fits, we can estimate an expansion time scale with the expression $t_{exp} = 1/\gamma \kappa$, where $\gamma = 1.0227$ $\rm{pc\ Myr^{-1}\ km^{-1}\ s}$ is a conversion factor to change the time-scale in Myr. This gets as results $t_{x,exp}=13.97^{+1.29}_{-1.10}$, $t_{y,exp}=24.44^{+1.96}_{-1.81}$ and $t_{z,exp}=48.89^{+12.12}_{-8.08}$ Myr. Additionally, we also found a correlation in the stars' speed with the radial distance, giving a slope of $\kappa_r = 0.094\pm 0.004$ $\rm{km\ s^{-1}\ pc^{-1}}$, and $t_{r,exp}=10.40^{+0.46}_{-0.43}$ Myr. Note that the expansion time in every profile is different, indicating a clear anisotropic expansion, probably caused by multiple mechanisms in the star formation history~\citep{cantat:2019}. However, several features must be taken into account. We obtained large error bars in the $x-v_x$ plane due to the strong dependency on $\varpi$. In the $y-v_y$ plane, we also see the presence of a prominent bulk of stars caused by the dense Orion Nebular Cloud (ONC) region, which is underestimating the likely real slope of the profile. If we remove the stars from the Orion A region (i.e. ONC and Ori-South), we get a time-scale of $t_{exp,y}=11.11^{+0.67}_{-0.60}$ Myr, which is much closer than the obtained in the $v_x-x$ profile.

Ultimately, for the $z-v_z$ plane, we point out that the expansion time is much larger than the estimated ages of the stellar populations found in Orion. Although this time-scale is anomalously big, we identified a bifurcation, as seen in the third panel of Figure~\ref{fig:Profile_Pos_Vel}. Therefore, by considering the stars located in both branches, we fit a linear model separately, as seen in Figure~\ref{fig:zw_bifurcation}. The first branch has a gradient of $\kappa_{z1}=0.145\pm 0.012$ $\rm{km\ s^{-1}\ pc^{-1}}$, represented by clusters Briceño-1A, Briceño-1B, $\sigma$ Ori and OBP-Near in red dots. This corresponds to a $t_{exp,z1}= 6.74^{+0.60}_{-0.51}$ Myr. The second branch shows a gradient of $\kappa_{z2}=0.035\pm 0.009$ $\rm{km\ s^{-1}\ pc^{-1}}$, which is formed by clusters Ori-North, Ori-East, OBP-Far, OBP-b and OBP-d in blue dots. In this case, the expansion time was calculated as $t_{exp,z2}= 27.87^{+9.65}_{-5.68}$ Myr. These results provide more reasonable time-scales, indicating that, on average, the populations from the first branch tend to have an earlier evolutionary stage than those in the second branch. Although the second branch is composed mainly of clusters belonging to the Orion D region, where reported ages are up to 21 Myr~\citep{kos:2020}, the dispersion of the sources causes that $t_{exp,z2}$ might be overestimated, considering the large upper quartile obtained ($\sim10$~Myr). The previous analysis gives weight to the hypothesis about the presence of a cavity formed by the relative movement of both populations in opposite directions, which could be related to the void found in~\citet{foley:2023}. To explain why the cavity is only visible in the $Z$ direction, we require a detailed dynamical analysis that is beyond the scope of this paper.

In summary, our results indicate that the Orion OB1 association generally exhibits a ballistic expansion, which means that further stars move much faster than those located near the centre of the complex. \citet{swiggum:2021} also conducted a similar analysis using \textit{Gaia} DR2, finding an expansion affecting only the OBP-Near and Briceño-1 clusters. No other evidence of expansion was identified in the remaining clusters recovered in their sample. Thus, our result indicates that the higher precision of the \textit{Gaia} DR3 data allows us to improve on that previous estimate.

\section{Small Structure Regime} \label{sec:ssr}

For a detailed study of the OSFC, we performed our second analysis using \textsc{hdbscan} in the search for smaller structures. From the \textit{main sample}, we recovered 40 clusters represented by 5,254 sources. By evaluating the spatial distribution of stars, six groups were discarded after being classified as extended, according to the approach applied in the \textit{Big Structure} regime (see Appendix~\ref{apendix_b:extended_clusters}). The remaining 34 groups were classified under two types: \textit{Small Groups}, defined as groups with stars not belonging to the \textit{Big Structures} found in the Section~\ref{sec:bsr}; and \textit{Sub-structures}, defined as groups whose stars are also members of a large structure.

After visual inspection, we identified 14 clusters established as small groups and 12 as sub-structures. The eight remaining groups correspond to the \textit{Big Structures}: Briceño-1A, Briceño-1B, Ori-East, OBP-far, $\sigma$ Orionis, OBP-b, OBP-d, and Orion Y. These groups preserve the same number of members in the small regime, indicating that they are not sensitive to the value of \texttt{min\_sample} within the range used in this work. In the following sections, we will discuss our findings on the two classifications mentioned earlier. To provide the reader with a clearer understanding, we separated the recovered clusters into those whose centroids are located in the $\lambda$ Orionis area ($\bar{\delta} > 4^\circ$) and those in the OB1 association ($\bar{\delta} < 4^\circ$).

\subsection{Recovery of small groups}
\label{sec:small_groups}

We detected a total amount of 14 \textit{small groups} in the OSFC. For the OB1 association, there are 11 groups: 7 previously found in the literature and 4 reported as candidates of new clusters. On the other hand, the remaining 3 groups are known clusters located in the surroundings of the $\lambda$ Orionis association. Complete information with the median value of the observable parameters can be found in Table~\ref{table:SS}.

{\renewcommand{\arraystretch}{1.24}
\begin{table*}
\centering
\caption{Median observational parameters of the small clusters recovered in the \textit{Small Structure} regime with the 16th and 84th percentiles. The first column is the identifier related to Figure~\ref{fig:cluster_results}. $N_T$ corresponds to the total number of star in the \textit{main sample} and with RV in parenthesis. The last column shows the references that have studied the clusters using \textit{Gaia} data. Median RV value is calculated if the cluster has at least six stars with available information.}
\label{table:SS}
\begin{tabular}{clcccccccl}
\hline
Cluster & \multicolumn{1}{c}{Name} & \begin{tabular}[c]{@{}c@{}}$\bar{\alpha}$\\ {(}deg{)}\end{tabular} & \begin{tabular}[c]{@{}c@{}}$\bar{\delta}$\\ {(}deg{)}\end{tabular} & \begin{tabular}[c]{@{}c@{}}$\bar{\varpi}$\\ {(}mas{)}\end{tabular} & \begin{tabular}[c]{@{}c@{}}$\bar{\mu}_\alpha^*$\\ {(}$\rm{mas}\ yr^{-1}${)}\end{tabular} & \begin{tabular}[c]{@{}c@{}}$\bar{\mu}_\delta$\\ {(}$\rm{mas}\ yr^{-1}${)}\end{tabular} & \begin{tabular}[c]{@{}c@{}}$\overline{RV}$\\ {(}$\rm{km}\ s^{-1}${)}\end{tabular} & \begin{tabular}[c]{@{}c@{}}$N_T$\\ (with RV)\end{tabular} & References \\ \hline\hline
a  & $\lambda$ Ori South A  & 85.96$^{+1.16}_{-1.77}$ & 6.54$^{+1.17}_{-1.94}$ & 2.29$^{+0.10}_{-0.08}$ & -2.99$^{+0.58}_{-0.48}$ & -1.47$^{+0.54}_{-0.32}$ & $-$ & 136 (1) & 2, 3  \\
b  & $\lambda$ Ori South B$^*$  & 79.41$^{+0.43}_{-0.98}$ & 7.39$^{+1.60}_{-0.37}$ & 2.64$^{+0.14}_{-0.13}$ & 2.33$^{+0.28}_{-0.30}$ & -3.66$^{+0.37}_{-0.24}$ & $-$  & 72 (3) & 2 \\
c  & L1562    & 75.93$^{+2.89}_{-0.36}$ & 12.85$^{+0.38}_{-2.82}$ & 3.26$^{+0.21}_{-0.13}$ & 1.67$^{+0.34}_{-0.29}$  & -3.50$^{+0.21}_{-0.38}$ & $-$ & 75 (0) & 2 \\
d  & L1622$^*$   & 87.52$^{+0.10}_{-0.09}$ & 2.82$^{+0.17}_{-0.33}$ & 2.53$^{+0.04}_{-0.10}$ & -1.13$^{+0.28}_{-0.19}$ & -1.83$^{+0.20}_{-0.14}$ & $-$ & 51 (0) & 1, 2  \\
e  & OBP-West       & 81.46$^{+0.48}_{-0.40}$ & 0.36$^{+1.03}_{-0.60}$ & 2.44$^{+0.05}_{-0.05}$ & 0.06$^{+0.30}_{-0.28}$ & 1.35$^{+0.23}_{-0.23}$ & 27.60$^{+0.70}_{-1.46}$ & 58 (19) & 2 \\
f  & L1616 A$^{\dagger}$        & 79.82$^{+0.67}_{-0.58}$ & -1.25$^{+0.71}_{-0.60}$ & 2.81$^{+0.07}_{-0.09}$ & 1.52$^{+0.10}_{-0.10}$ & -0.46$^{+0.13}_{-0.18}$ & 20.92$^{+1.23}_{-0.83}$ & 83 (9) & 2, 3 \\
g  & L1616 B$^{\dagger}$       & 77.21$^{+0.53}_{-0.39}$ & -2.96$^{+0.46}_{-0.38}$ & 2.65$^{+0.05}_{-0.06}$ & 1.22$^{+0.19}_{-0.21}$  & -0.78$^{+0.25}_{-0.38}$ & 22.54$^{+1.65}_{-0.42}$ & 67 (7) & 2, 3 \\
h  & L1634      & 80.82$^{+0.48}_{-0.40}$ & -4.27$^{+0.54}_{-0.49}$ & 2.68$^{+0.08}_{-0.04}$ & 1.85$^{+0.21}_{-0.20}$ & -1.02$^{+0.24}_{-0.18}$ & $-$ & 88 (1) & 2, 3 \\
i  & Rigel        & 79.03$^{+0.59}_{-0.93}$ & -6.57$^{+1.25}_{-1.58}$ & 3.46$^{+0.09}_{-0.18}$ & 1.72$^{+0.27}_{-0.34}$ & -2.90$^{+0.28}_{-0.230}$ & $-$ & 67 (0) & 2, 3  \\
j  & Orion A East         & 87.26$^{+1.00}_{-0.59}$ & -5.27$^{+0.62}_{-0.57}$ & 2.30$^{+0.09}_{-0.06}$ & -2.45$^{+0.16}_{-0.27}$ & 0.31$^{+0.14}_{-0.26}$ & $-$ & 54 (1) & 2, 3  \\
k  & OBP-Centre$^{\ddagger}$    & 84.70$^{+0.67}_{-0.43}$  & -0.65$^{+0.71}_{-1.22}$ & 2.50$^{+0.07}_{-0.05}$  & -0.03$^{+0.24}_{-0.40}$ & -0.61$^{+0.24}_{-0.55}$ & 28.05$^{+2.76}_{-2.59}$ & 79 (19) & --  \\
l  & OBP-South$^{\ddagger}$  & 82.29$^{+0.56}_{-0.97}$ & -3.08$^{+0.83}_{-1.27}$ & 3.01$^{+0.03}_{-0.06}$ & 1.10$^{+0.15}_{-0.09}$  & -0.25$^{+0.07}_{-0.10}$ & 20.38$^{+9.25}_{-0.69}$ & 51 (7) & --  \\
m  & Ori Far West$^{\ddagger}$  & 77.85$^{+0.83}_{-0.53}$ & -4.09$^{+1.11}_{-1.40}$ & 2.93$^{+0.10}_{-0.08}$ & 0.34$^{+0.22}_{-0.12}$ & -1.29$^{+0.28}_{-0.28}$ & 18.80$^{+2.88}_{-3.31}$ & 68 (6) & --  \\
n  & OBP-East$^{\ddagger}$  & 86.81$^{+1.28}_{-1.10}$ & -1.35$^{+1.83}_{-1.84}$ & 2.19$^{+0.10}_{-0.13}$  & 2.21$^{+1.00}_{-1.73}$ & -4.11$^{+0.46}_{-0.54}$ & 18.62$^{+0.89}_{-0.69}$ & 91 (6) & -- \\ \hline
\multicolumn{10}{l}{$^*$ Clusters recovered by~\citet{chen:2020} only with one clustering algorithm.}\\
\multicolumn{10}{l}{$^{\dagger}$ Clusters recovered by~\citet{chen:2020} as one only group.}\\
\multicolumn{10}{l}{$^{\ddagger}$ Clusters reported as possible new discoveries.}\\
\multicolumn{10}{l}{\textbf{References:} (1)~\citet{kounkel:2018_apogee}, (2)~\citet{chen:2020}, (3)~\citet{swiggum:2021}.}\\
\end{tabular}
\end{table*}}

\begin{figure*}
    \centering
    \includegraphics[width=0.95\textwidth]{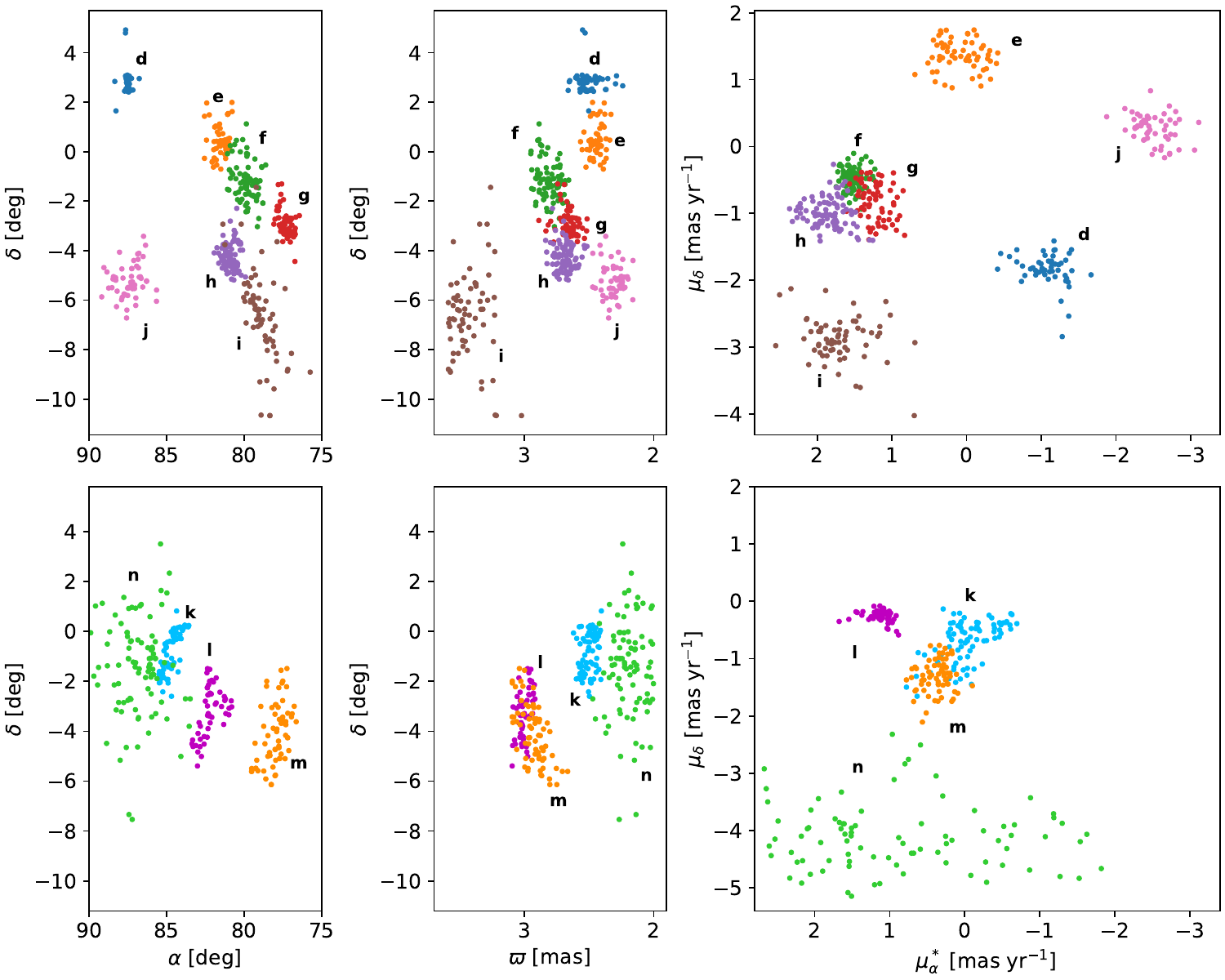}
    \caption{Small structures recovered in the OB1 association and their distribution in the sky plane (left panel), parallax versus declination (middle panel), and proper motions (right panel). In the first row, from a to j, we plotted the known clusters already reported in the literature; in the second row, we show, from k to n, the candidates as new clusters. Literals are related to the astrometric information given in table~\ref{table:SS}.}
    \label{fig:cluster_results}
\end{figure*}

\subsubsection{Known small clusters in the OB1 association}

In the top panels of Figure~\ref{fig:cluster_results}, we show the distribution of the seven known small clusters in terms of sky coordinates, parallax and the vector--point diagram. Unlike the \textit{Big Structures}, which seem to be agglomerated in a central region, these small clusters appear spatially sparse in the outskirts of the OB1 association.

Towards the East sector, we have the clusters L1622 and Orion A East (clusters a and j) as the smallest ones, with 51 and 54 stars, respectively. The former belongs to the classical Orion B structure found by~\citet{kounkel:2018_apogee}, sharing a similar proper motion behaviour as the Ori-East group detected in the \textit{Big Structure} regime. On the other hand, the Orion A East group is located towards the Southern area with a wider dispersion in the sky plane and a median distance of 443$\pm$16 pc, which is the furthest among the known OB1 association clusters. 

On the Northern sector, we have OBP-West (cluster e) with 58 stars reported by~\citet{chen:2020} at an approximate distance of $409\pm 8$ pc. Its position indicates that this stellar group is close to Ori-North, with a separation of $\sim 20$ pc between their centroids. Moreover, the vector--point diagram shows that the motion of the stars diverges along the $\mu_{\alpha}^*$ axis, suggesting that OBP-West might be formed of two smaller groups with distinct kinematics.

On the West side, we have L1616 A and L1616 B (clusters f and g) with 88 and 83 stars, respectively, which were initially identified as a single cluster by~\citet{chen:2020} using \textit{Gaia} DR2 data. Now, with the enhanced precision provided by \textit{Gaia} DR3, we can detect a clear separation into two individual distributions. Additionally, with L1634 (cluster h) located slightly further south, these groups share similar $\varpi$ and $\mu$, suggesting that L1616 A, L1616 B and L1634 likely originated from the same molecular cloud filament. Finally, the Rigel group (cluster i), with 68 stars, is the closest cluster to the Sun, at a distance of approximately 293$\pm$11 pc (comparable to that found for Orion Y in the \textit{Big Structure} regime). The Rigel group also exhibits a very isolated proper motion distribution, suggesting that its formation may have occurred under different conditions compared to the other clusters in the region.

\subsubsection{Candidates to new clusters in the OB1 association}
\label{sec:small_new}

\begin{figure*}
    \centering
    \includegraphics[width=0.94\textwidth]{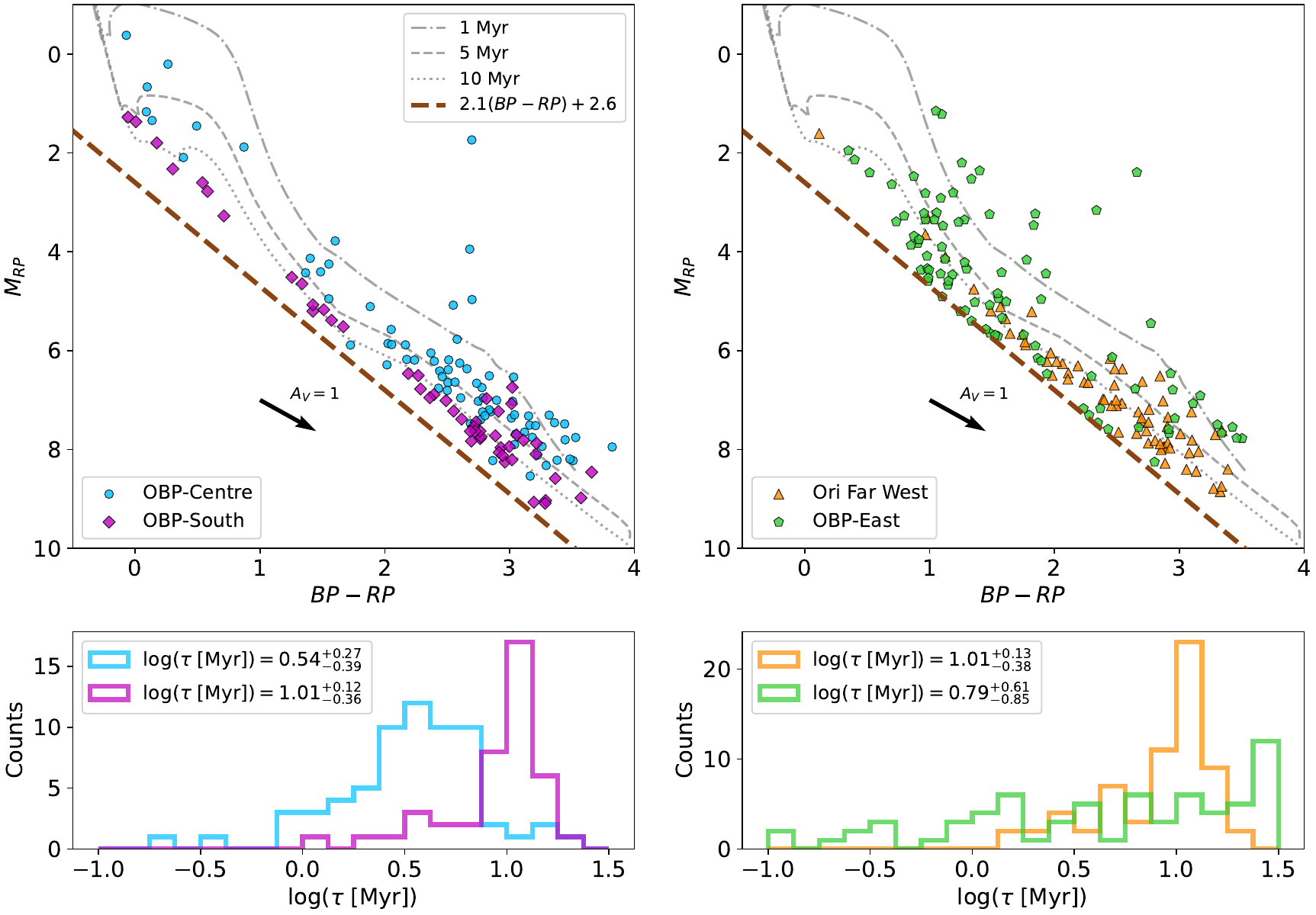}
    \caption{\textit{Top panels}: distribution of the four candidates to new clusters: OBP-Centre, OBP-South, Ori Far West and OBP-East, in a CMD employing \textit{Gaia} DR3 photometry. \textit{Bottom panels}: histograms with the estimation of $\log(\tau)$ inferred from equation~\ref{eq:age}. Values shown are the median with the 16th and 84th percentiles. The estimated ages for the clusters are: $3.45^{+2.47}_{-1.86}$ (OBP-Centre), $10.33^{+2.31}_{-1.33}$ (OBP-South), $10.24^{+2.42}_{-1.34}$ (Ori Far West), and $6.19^{+7.04}_{-4.10}$ Myr (OBP-East).}
    \label{fig:new_isochrones}
\end{figure*}

In this section, we describe the groups labelled as OBP-Centre, OBP-South, Ori Far West and OBP-East (clusters k, l, m and n), classified as candidates to new clusters. We show the distribution of these groups in the sky projection, parallax and vector--point diagram in the bottom panels of Figure~\ref{fig:cluster_results}.

Immersed within the OBP region, we obtained OBP-Centre at 408$\pm$10 pc with 79 stars. It has a median proper motion of $\mu_\alpha^* \sim -$0.03 and $\mu_\delta \sim -$0.61 $\rm{mas}\ s^{-1}$, indicating a distinct kinematics compared to its closest neighbours in the OBP area. Its spatial distribution in the sky projection shows an arched shape surrounding some of the largest OBP clusters. Moving further south, we found the second cluster candidate, OBP-South, with 51 stars at $\sim 332.2\pm 5.5$ pc. This distance is comparable with Ori Far West at $341.3\pm11.5$ pc, the third cluster candidate with 68 stars situated above the Rigel group in the sky projection and predominant kinematics along the $\delta$ axis.

Finally, we have the fourth cluster, OBP-East, the furthest group in the \textit{Small Structure} regime at a median distance of $452.6\pm 20.45$ pc. Even though it was classified as a cluster for having its spatial distribution inside the $5\sigma$ limit in the SIR$_{\alpha}$ versus SIR$_{\delta}$ and SIR$_{\varpi}$ versus SIR$_{\delta}$ projections (see Appendix~\ref{apendix_b:extended_clusters} for more details), OBP-East exhibits a particular spread in the $\mu_{\alpha*}$ axis alongside the highest proper motion in the $|\mu_\delta|$component. Thus, we suggest caution when using this group as a real cluster.
 
To further analyse the nature of these populations, we explored some physical parameters through their photometry. For example, in Figure~\ref{fig:new_isochrones} (top panels), we plotted the four candidates on the $BP-RP$ colour--magnitude diagram (CMD), including the 1, 5 and 10-Myr \textsc{parsec} isochrones alongside the 30-Myr linear approximation shown in Figure~\ref{fig:CMD_cut}. We defined an age calibrator using the interval $1.0 < BP-RP < 3.5$ since the extinction vector is approximately parallel to the theoretical isochrones in the CMD. First, we computed the difference between the theoretical $RP$ magnitude from the 1, 3, 5, 10, 20, and 50-Myr \textsc{parsec} isochrones and the $RP$ reference magnitude ($RP_{REF}$), derived by interpolating the corresponding $BP-RP$ colour in the 30-Myr linear approximation. When $\Delta RP$ is compared with the isochrone ages in a logarithmic scale (log($\tau\ [\text{Myr}]$)), we found the following linear expression:%
\begin{equation}
    \label{eq:age}
    \log(\tau\ [\text{Myr}]) = 1.499 (\pm 0.012) - 0.665(\pm 0.010)\times\Delta RP.
\end{equation}

We estimated the age of each source from equation~\eqref{eq:age}, where $\Delta RP$ represents the difference between its absolute $RP$ magnitude and the corresponding $RP_{REF}$. Afterwards, we obtained a distribution that can assign a statistical age to every cluster, specifically, the median within the 16th and 84th percentiles. Here, it is important to mention that equation~\ref{eq:age} represents an approximate method for comparing age estimation between several young stellar clusters based only on the \textsc{parsec} evolutionary model.

In Figure~\ref{fig:new_isochrones} (bottom panels), we show the age distributions of the stars that belong to the four clusters. After measuring the median, we estimated that OBP-Centre seems to be the youngest cluster with $\tau = 3.45^{+2.47}_{-1.86}$ Myr. This result agrees that OBP-Centre is located in a region where clusters are highly embedded in gas. On the other hand, since OBP-East has a high dispersion in the CMD, we obtained an uncertain age of $\tau = 6.19^{+7.04}_{-4.10}$ Myr. This result, alongside the scattering in the $\mu^*_\alpha$ component in Figure~\ref{fig:cluster_results}, suggests that OBP-East might include contamination or several non-coeval minor populations. Finally, the remaining groups Ori Far-West and OBP-South show similar ages of $\tau = 10.33^{+2.31}_{-1.33}$ and $\tau = 10.24^{+2.41}_{-1.34}$ Myr, respectively. Since they are located on the west side of the complex, the absence of gas in their surroundings suggests that these small groups are in a late evolutionary state compared to OBP-Centre.

\subsubsection{Known small clusters in $\lambda$ Orionis}
\begin{figure*}
    \centering
    \includegraphics[width=0.32\textwidth]{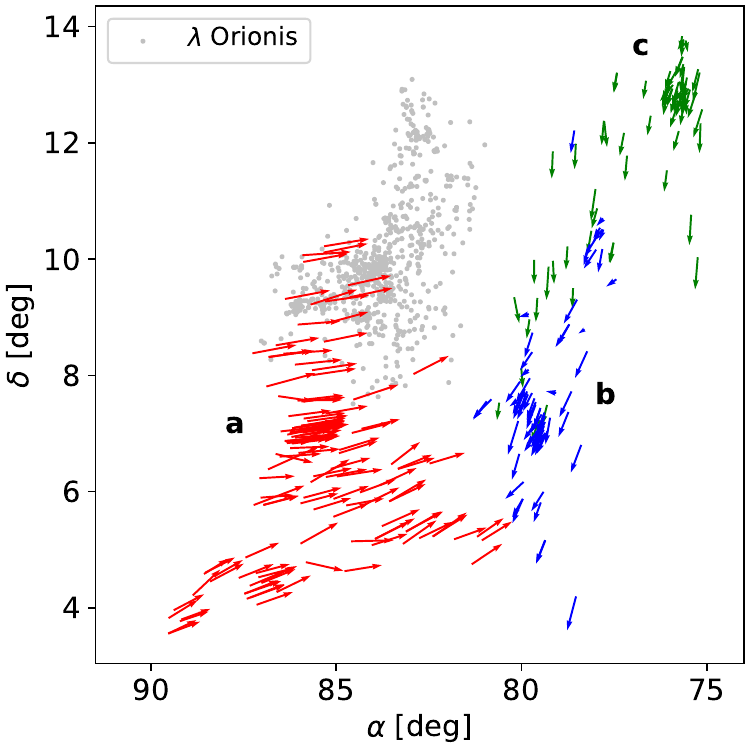}
    \includegraphics[width=0.32\textwidth]{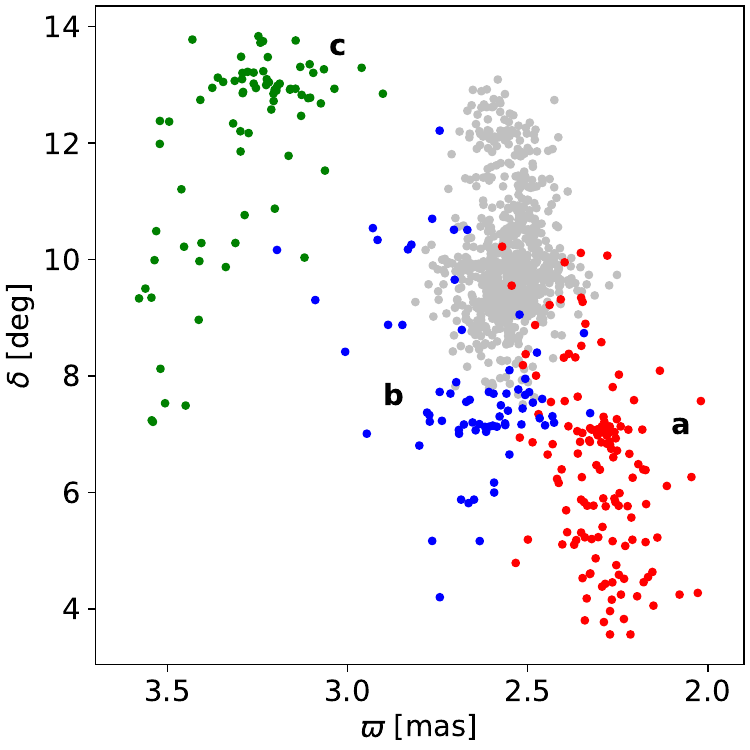}
    \includegraphics[width=0.32\textwidth]{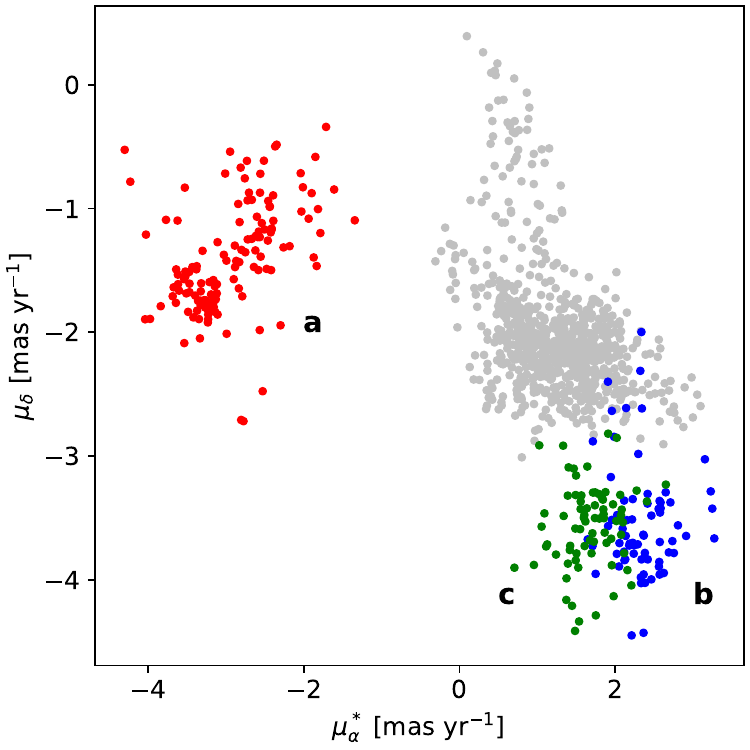}
    \caption{Small groups recovered in the surroundings of the $\lambda$ Orionis association in terms of Equatorial Coordinates (left panel), parallax versus declination (middle panel), and vector--point diagram (right panel). In the sky--plane distribution, proper motion vectors are included after subtracting the median value of the large structure ($\lambda$ Orionis) in order to see their kinematics. Additionally, in all panels $\lambda$ Orionis (cluster 1 in Table~\ref{table:BS}) is included for reference.}
    \label{fig:small_groups_lamori}
\end{figure*}

\textsc{hdbscan} detected three \textit{small groups} in the surroundings of $\lambda$ Orionis. Figure~\ref{fig:small_groups_lamori} presents the spatial distribution of these clusters in the sky projection (including proper motion vectors), declination versus parallax and vector--point diagram. Below $\lambda$ Orionis, we found the groups $\lambda$ Ori South A (cluster a) and $\lambda$ Ori South B (cluster b) with 136 stars (which is very close to the \texttt{min\_sample} threshold used in the \textit{Big Structure} regime) and 72 stars respectively, both previously mentioned by~\citet{chen:2020}. These groups show certain proximity, locating $\lambda$ Ori South A at 436$\pm$11 pc and $\lambda$ Ori South B at 379$\pm$8 pc away. This leaves just a difference of $\sim$56 pc between both centroids. Regarding kinematics, $\lambda$ Ori South A demonstrates a relative movement towards the region's West side, suggesting a possible future encounter with $\lambda$ Ori South B, which is moving mainly to the south. 

The last cluster is L1562 (cluster c) with 74 stars, also reported by~\citealp{chen:2020}, mostly agglomerated at $\alpha = 75.93^{\circ}$ and $\delta=12.85^{\circ}$. The vector--point diagram in figure~\ref{fig:small_groups_lamori} shows that L1562 has similar kinematics with $\lambda$ Ori South B. However, they differ in distance, finding L1562 at 307$\pm$6 pc. Even though these three small groups display particular kinematic behaviours, the lack of RV data prevents a 6D phase space analysis.

\subsection{Identification of sub-structure}
\label{sec:sub_structures}

\textsc{hdbscan} also identified that five major groups exhibit clear separation in smaller distributions with kinematics associated with their original \textit{Big Structure}. Four belong to the OB1 association, whereas the last corresponds to $\lambda$ Orionis. The remaining eight large groups in Table~\ref{table:BS} did not show sub-structures with \texttt{min\_sample}=50.

\subsubsection{Sub-structure in the OB1 association}

\begin{figure*}
    \centering
    \includegraphics[width=0.95\textwidth]{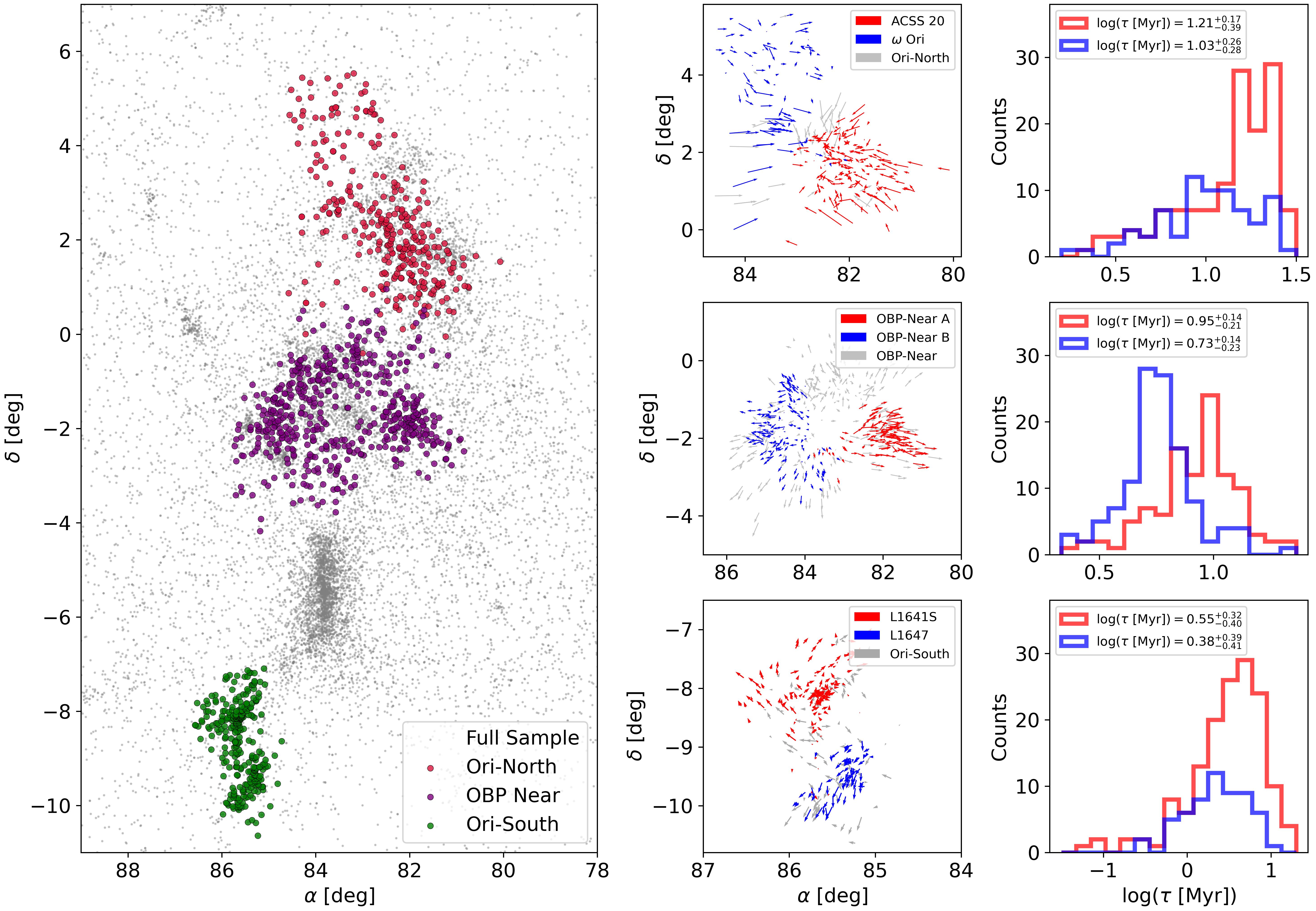}
    \caption{Sub-structure found in the OB1 association from large clusters: Ori-North, OBP-Near and Ori-South. \textit{Left panel}: general distribution of the astrometric sample showing the three big groups with clear sub-structure. \textit{Middle panels}: minor clusters recovered with their distribution in the sky plane and proper motion vector after subtracting the median value of the large group. \textit{Right panels}: distributions of $\log (\tau)$ of every sub-structure including the median value and the 16th and 84th percentile. These estimate the following ages: $16.07^{+1.48}_{-2.46}$ (ASCC 20), $10.65^{+1.82}_{-1.92}$ ($\omega$ Ori), $8.95^{+1.37}_{-1.63}$ (OBP-Near A), $5.41^{+1.37}_{-1.68}$ (OBP-Near B), $3.51^{+2.09}_{-2.50}$ (L1641S), and $2.38^{+2.44}_{-2.56}$ Myr (L1647).}
    \label{fig:substructure_results}
\end{figure*} 

\begin{figure*}
    \centering
    \includegraphics[width=0.95\textwidth]{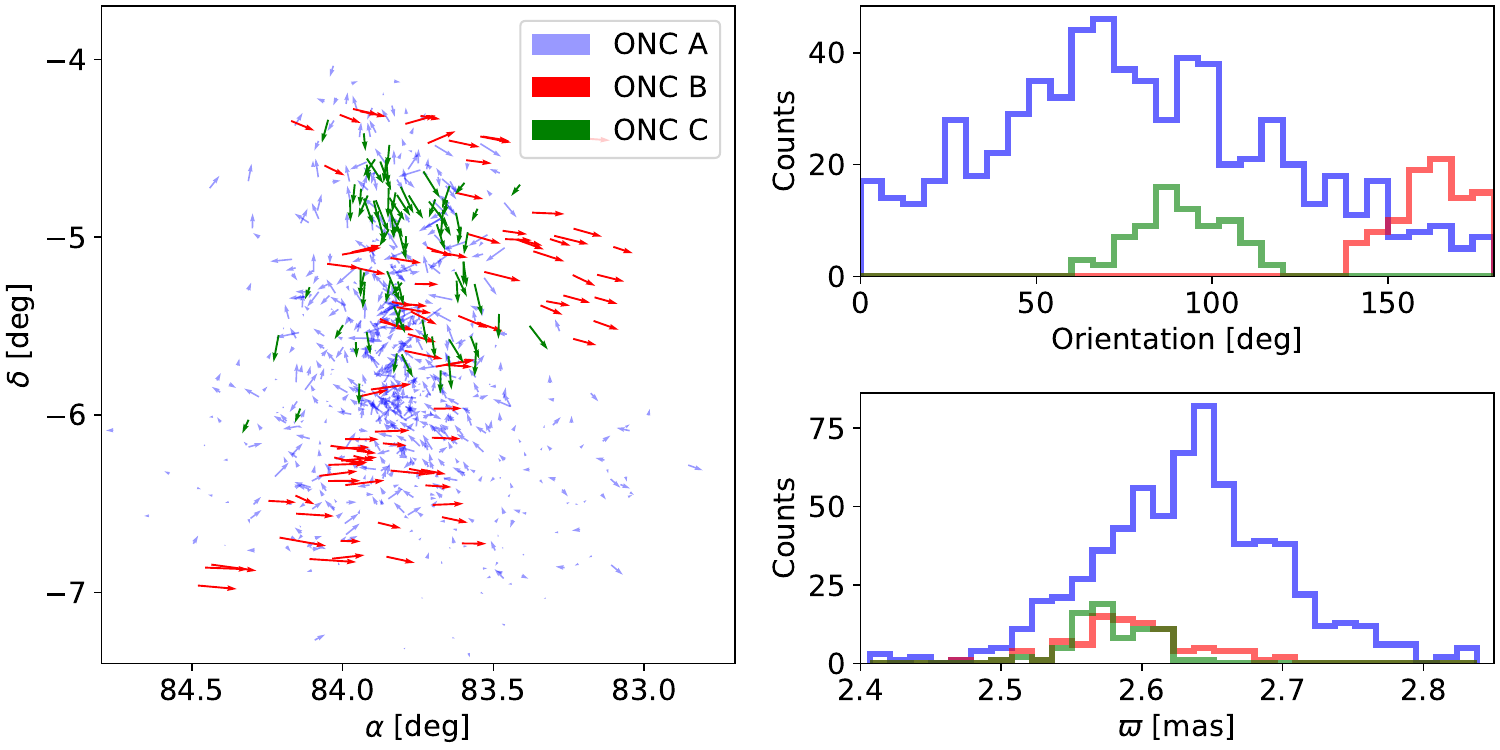}
    \caption{Kinematic behaviour of the clusters identified in the ONC region. \textit{Left panel}: spatial distribution in the sky plane with proper motion vectors of the three sub-structures recovered after subtracting the median value of the large group. \textit{Top-right panel}: histogram showing the orientation of the proper motion vectors in each group with respect to the right ascension axis. \textit{Bottom-right panel}: histogram of parallax for the three sub-structures.}
    \label{fig:onc_result}
\end{figure*}

For the Orion OB1 association, presented in Figure~\ref{fig:substructure_results}, we can see the sub-structure in the groups Ori-North (cluster 2), OBP-Near (cluster 10), and Ori-South (cluster 12), comparing the distribution of stars in the sky projection (left panel), the proper motion vectors after subtracting the median value of the main groups (middle panels), and the age estimation using the \textit{Gaia} photometry (right panels).

On the Northern side, where Ori-North is located, we found a division into two small groups: $\omega$ Ori and ASCC 20, also recovered by~\citet{chen:2020}. The orientation of the proper motions shows that stars follow apparent visual contraction. Simultaneously, in $\omega$ Ori (red vectors), we can identify that the stars below $\delta\approx 3^{\circ}$ have a different kinematic with respect to the rest of the group, with proper motion components following an eastbound orientation. This might suggest the presence of a smaller group embedded in the main sub-structure. However, a deeper analysis will be considered in detail for future work. Moreover, if we apply equation~\ref{eq:age}, we can estimate the likely age of the sub-structures. After calculating the median of the age distribution, we observed that ASCC 20 has a value of $\tau = 16.07^{+1.48}_{-2.46}$ Myr, which seems to be a more evolved cluster than $\omega$ Orionis with $\tau = 10.65^{+1.82}_{-1.92}$ Myr. 

On the central region, OBP-Near shows a ring-like shape with a visual radial expansion of its stars when the median of $\mu_\alpha^*$ and $\mu_\delta$ is subtracted from the data. We can estimate the expansion time of this cluster with the expression $t_{exp}=1/\gamma \kappa$ in a position-velocity profile, using the total velocity and distance to the centroid, obtaining $\sim$7.52 Myr, which is within 1-$\sigma$ according with the age estimation reported by~\citet{swiggum:2021} of 6.2$\boldsymbol{^{+3.5}_{-2.4}}$ Myr. After applying \textsc{hdbscan}, we detected two sub-structures named OBP-Near A and OBP-Near B for identification. Both sub-structures are moving in opposite directions along the right ascension axis. At the same time, in terms of age estimation, we calculated $\tau = 8.95^{+1.37}_{-1.63}$ Myr for OBP-near A and $\tau = 5.41^{+1.37}_{-1.68}$ Myr for OBP-near B. This result indicates that the West side of the main cluster is more evolved than the East side.

In Ori-South, \textsc{hdbscan} split the structure into two smaller groups: L1641S and L1647. This region has a peculiar morphology, suggesting ``twisting" motions in its kinematics. However, more information is needed to give a conclusion about this behaviour. When their photometry is analysed, we obtained close estimated ages with values $\tau = 3.51^{+2.09}_{-2.50}$ Myr for L1641S and $\tau = 2.38^{+2.44}_{-2.56}$ Myr for L1647, which would place them as the less evolved structures in our analysis. This statement is reasonable, considering that Ori-South is located in the tail of the Orion A molecular cloud, meaning these clusters could still be embedded in gas.

One of the regions not included in Figure~\ref{fig:substructure_results} but also divided into sub-structures was the ONC. However, due to its complex distribution and high star density, the associated kinematics must be analysed cautiously. In general, the kinematics of the stars are highly intricate and stochastic for finding a characteristic pattern in the proper motion space. Nevertheless, the clustering algorithm in the \textit{Small Structure} regime under the 5D parameter space showed three different groups named ONC A, ONC B, and ONC C. In Figure~\ref{fig:onc_result}, we show some spatial and kinematical features of the groups mentioned above in terms of their distribution on the sky position, proper motion orientation,\footnote{\textit{Orientation} is defined as the angular position of the proper motion vectors projected in the sky with respect to the right ascension axis.} and parallax.

The first group (ONC A) is the densest distribution with 676 stars. The remaining clusters (ONC B and ONC C) are much smaller in star number, with 93 and 76 stars, respectively. When the median proper motion of the main group is subtracted, as seen in the left panel of Figure~\ref{fig:onc_result}, it is evident that ONC B has an apparent motion from East to West and ONC C from North to South. While the smaller groups exhibit different kinematic behaviour than ONC A, based on the right panels in Figure~\ref{fig:onc_result}, we noticed that ONC B and ONC C are clusters with different proper motion orientations but similar distances. This result might indicate the presence of three independent stellar populations. Other works also find different stellar populations in the ONC region~\citep[e.g., ][]{beccari:2017}. However, more data are required to relate our sub-structures to the reported populations and validate a suitable scenario for the formation of this region.

\subsubsection{Sub-structure in $\lambda$ Orionis}
\label{subsection:lambda_ori}

\begin{figure}
    \centering
    \includegraphics[width=0.48\textwidth]{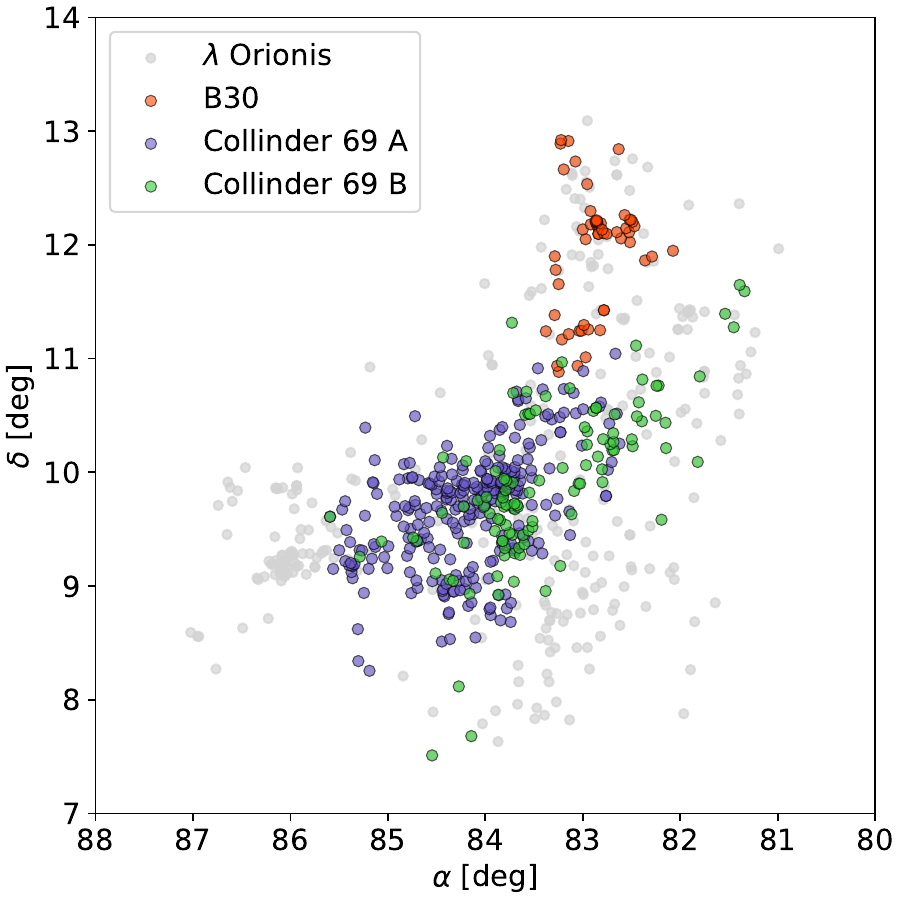}
    \caption{Distribution of the sub-structure found in the $\lambda$ Orionis cluster after applying \textsc{hdbscan} in the \textit{Small Structure} regime. In grey, we included the stars belonging to the large structure but with no membership to any sub-structure.}
    \label{fig:lambda_ori_substructure}
\end{figure}

\begin{figure*}
    \centering
    \includegraphics[width=0.96\textwidth]{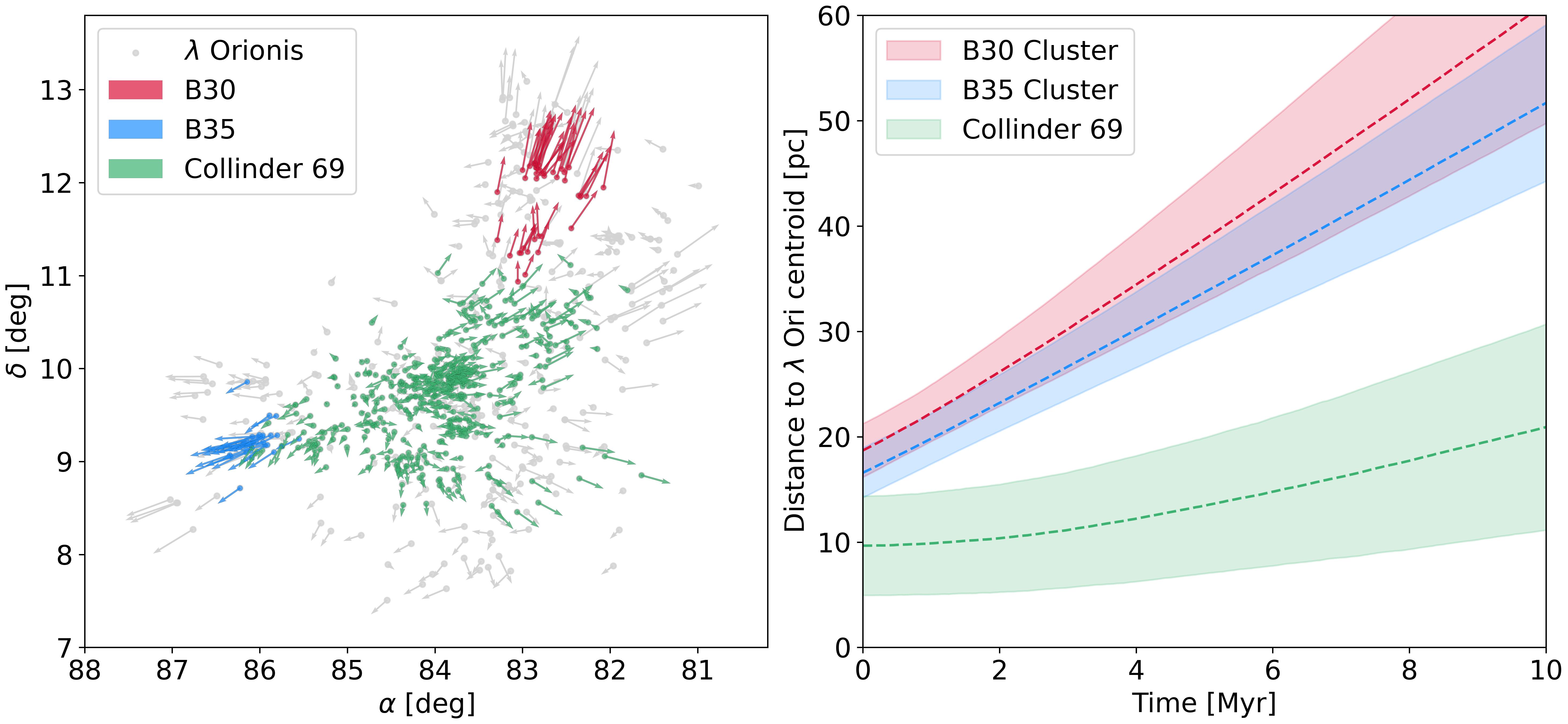}
    \caption{\textit{Left panel}: distribution of the recovered $\lambda$ Orionis sub-structures in the sky plane. Proper motion vectors are included with the subtraction of the median value obtained in table~\ref{table:BS}. In grey we included the stars that belong to the large structure but with no membership in any sub-structure. \textit{Right panel}: kinematic extrapolation of sub-structure positions with respect to the centroid of the main cluster at present time, shadowed regions are the $1\sigma$ error bars calculated from the Monte Carlo sampling.}
    \label{fig:lam-ori-substructure1}
\end{figure*}

In the $\lambda$ Orionis region, we found three sub-structures from the main group. Only one was identified as a known cluster: the group B30 located in the northern region as seen in Figure~\ref{fig:lambda_ori_substructure}. The other two clusters reported in the literature~\citep[e.g.][]{barrado:2007, mathieu:2008, kounkel:2018_apogee}: Collinder 69 and B35, do not clearly appear. Instead, the central region was divided into two agglomerations of stars that share the same distance ($\sim$398 pc). This means that the algorithm was not able to recover the known clusters when the \textit{main sample} is used at \texttt{min\_sample}=50, also indicating that the number of stars for B35 in the \textit{main sample} should be below 50 stars. When we tested \textsc{hdbscan} for \texttt{min\_sample} $<50$, it was not possible to recover the known stellar groups. On the contrary, the central region showed some minor sub-structures alongside extended groups. This result indicates a limitation in the algorithm for recovering the known clusters that compose the $\lambda$ Orionis association. 

As a solution, we applied our methodology using only the stars from the main group as  input for \textsc{hdbscan} with \texttt{min\_sample}=30. At this level, B30, B35, and Collinder 69 were obtained. The three recovered sub-structures are seen in the left panel of Figure \ref{fig:lam-ori-substructure1}. Most vectors show an outward direction when the median value of the proper motions from $\lambda$ Orionis is subtracted. The central cluster corresponds to Collinder 69 with 395 stars, where the ballistic expansion originates. The other two are the small groups B30 and B35 located at the outskirts of the main cluster~\citep{barrado:2007, mathieu:2008} with 43 and 30 stars, respectively. B30 and B35 formed in two minor clouds with an average age between 2 and 3 Myr, which is younger than the central part around $\sim$5 Myr~\citep{hernandez:2010}.

Kinematically, we can analyse the future position of the spatial centroids for the sub-structures previously detected. By repeating the same procedure followed in the \textit{Big Structure} regime, we evolved the whole system for the next 10 Myr. In the right panel of Figure~\ref{fig:lam-ori-substructure1}, we see the change in the relative position of each cluster concerning the current $\lambda$ Orionis centroid, showing that peripheral groups (B30 and B35) are moving away much faster than Collinder 69, which agrees with the ballistic expansion \citep{kounkel:2018_apogee}. Here, error bars for each group are calculated using Monte Carlo sampling from the 6D phase space.

\section{Likely Close Encounters in the OB1 association} \label{sec:merging}

\begin{figure*}
    \centering
    \includegraphics[width=0.95\textwidth]{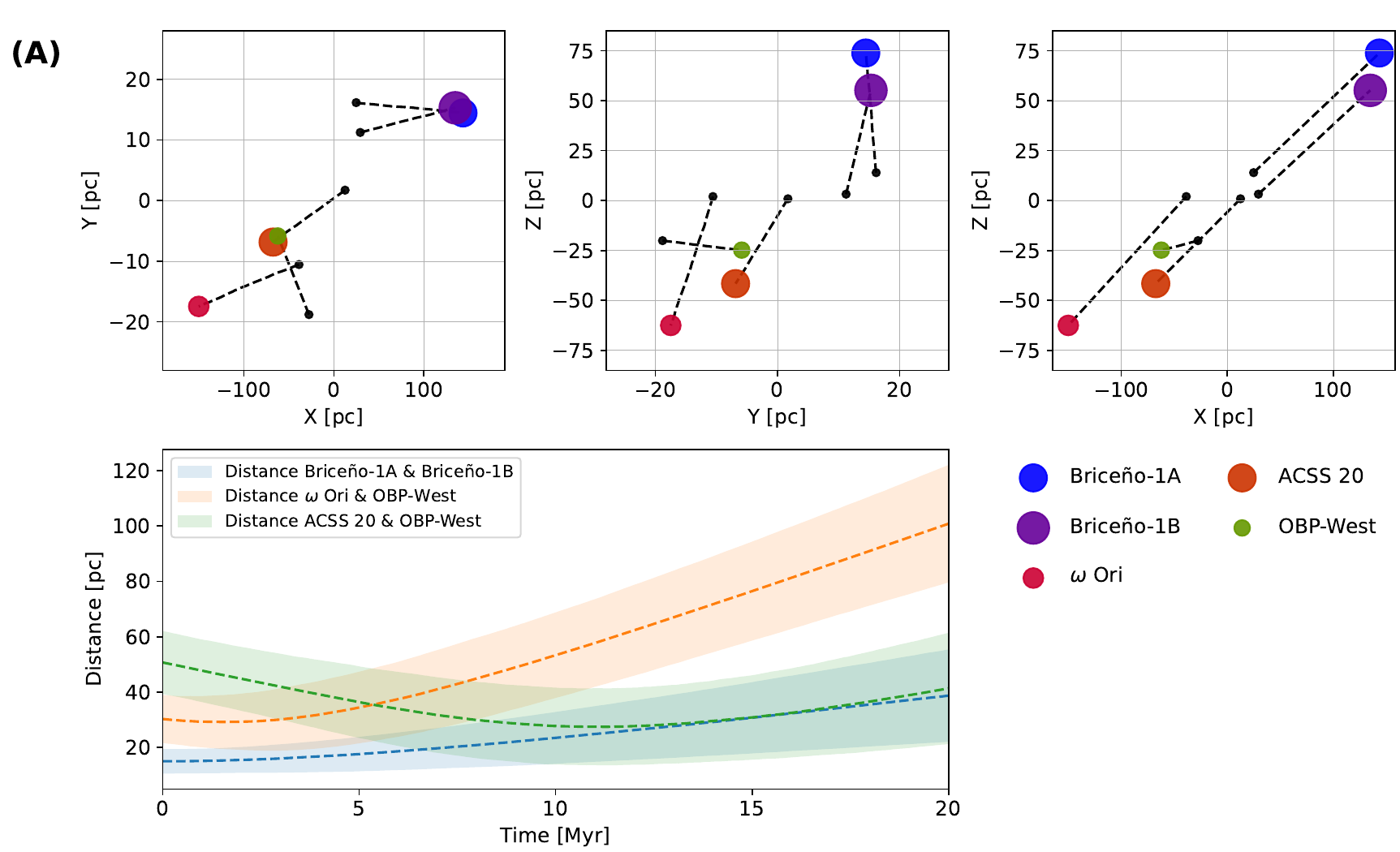}
    \hfill
    \includegraphics[width=0.95\textwidth]{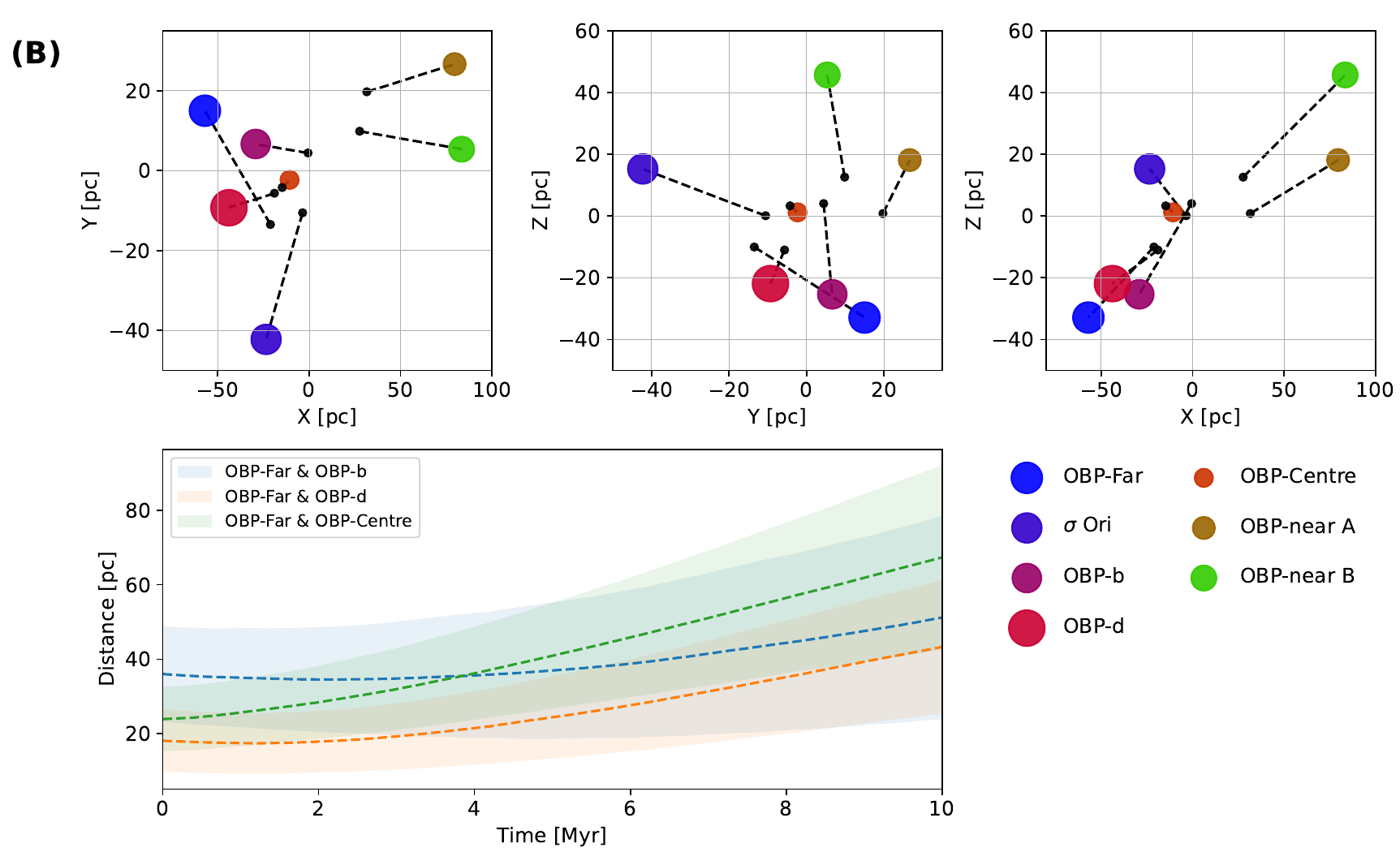}
    \caption{Simple simulation in the evolution of clusters in the OB1a association after 20 Myr (A) and the OBP region after 10 Myr (B). In both cases, the Cartesian projections show the kinematic extrapolation of the centroids assuming rectilinear motion (dashed lines). Bottom panels in both regions show the separation between particular clusters whereas the shadowed regions are the error propagation after applying Monte Carlo sampling.}
    \label{fig:cluster_mergings}
\end{figure*}

Up to this point, we were able to identify clusters in the \textit{Small Structure} regime. Nevertheless, further kinematic analysis is necessary to assess whether the identified structures will undergo gravitational interactions. In this section, making use of the RV information available, we carried out some simple simulations of rectilinear motion (similar to the one performed in the~\textit{Big Structure} regime) in those regions where  the cluster population is prominent and RV data from high-resolution spectroscopic surveys are available.

In the Northern side of the Orion OB1 association (also known as OB1a association), we evolved the position of the five residing clusters found in the \textit{Small Structure} regime: Briceño-1A, Briceño-1B, $\omega$ Orionis, ASCC20 and OBP-West, for the next 20 Myr, as shown in Figure~\ref{fig:cluster_mergings} (case A). As initial conditions, we calculated the centroids and the velocity vectors for each group, using the median of their respective distributions, with the size related to the number of stars (invariant along the simulation). The kinematic extrapolation shows that clusters belonging to Orion C (Briceño-1A and Briceño-1B) are moving away from those in Orion D ($\omega$ Orionis, ASCC20 and OBP-West). This behaviour is further supported by RV data, which reveals a difference of approximately $\sim 9.25$ km s$^{-1}$ between the two regions, consistent with the findings of~\citet{kounkel:2018_apogee}.

To determine close encounters, we calculated the distance between spatial centroids to evaluate their projection over time, as seen in Figure~\ref{fig:cluster_mergings} (top panels). We obtained that Briceño-1A and Briceño-1B showed a trend of distancing between them; on the other hand, ASCC20 and  OBP-West appear to have a highly perpendicular trajectory, which calls for a future close encounter. If we observe the distance of their centroids, the closest approach may occur in $\sim$11.37 Myr with a separation around 27.98$\pm$14.17 pc. Additionally, a possible close encounter between $\omega$ Orionis and OBP-West is expected in $\sim$1.30 Myr at 29.26$\pm$9.72 pc between their centroids. Any other combination of clusters always suggested a permanent separation.

We also analysed the OBP region, which is considered the richest environment in terms of cluster population. As it was proposed by~\citet{kounkel:2020_supernova}, \citet{swiggum:2021}, and \citet{foley:2023}, the Orion Belt could be the epicentre of an expansion process, probably triggered by a supernova event. This behaviour is shown in Figure~\ref{fig:cluster_mergings} (case B) using the 3D spatial evolution, where most groups clearly show a radial motion if we perform a rectilinear motion for the next 10 Myr. However, our analysis also indicates that the trajectory followed by OBP-far, the only cluster with no radial motion, has an approximation to OBP-b and OBP-d, expecting to reach the closest encounters in 2.43 and 1.03 Myr with centroid separation at 34.32$\pm$14.53 and 17.28$\pm$8.11 pc, respectively, as seen in Figure~\ref{fig:cluster_mergings} (bottom panel).

\section{Conclusions}\label{sec:conclusions}

The OSFC is a valuable laboratory in the study of star formation and evolution of YSCs as data provide a significant variety of kinematics and spatial distribution scenarios. In this work, we have analysed the YSCs in the OSFC under two regimes: the \textit{Big Structures} and the \textit{Small Structures}, based on the recovery of groups when a clustering algorithm is applied with different \texttt{min\_sample} values in a 5D space, constructed from the \textit{Gaia}'s observable parameters ($\alpha, \delta, \mu_\alpha^*, \mu_\delta, \varpi$). Also, we analysed the kinematic projection of the clusters with RV data to detect expansion processes and likely scenarios of close interactions in the Cartesian phase space. Here, we summarize the most important results.

\subsection{Big Structure regime}

13 groups were recovered along the complex covering the five main regions proposed by~\citet{kounkel:2018_apogee}: Orion A, B, C, D and $\lambda$ Orionis. These groups have a general distribution in a filamentary structure along the declination axis. When rectilinear motion is assumed for the clusters in the next 10 Myr (where the effect of Galaxy potential can be neglected), an anisotropic expansion process is detected in the 3D Cartesian space, with most of them following a radial displacement in the $X-Y$ projection. The $X-Z$ and $Y-Z$ projections show that clusters seem to move in two preferential directions.

From position-velocity profiles, we detect positive correlations in the data manifesting the expansion process. However, the gradients indicate that the OB1 association follows an anisotropic evolution. Additionally, in the $z-v_z$ projection, we detect a bifurcation, suggesting the presence of a cavity. The positive correlation in the position-velocities diagram could support the hypothesis of a supernova event in the central part of the OB1 association~\citep{kounkel:2020_supernova}, which is also linked with a devoid of gas between Orion C and Orion D~\citep{foley:2023}.

\subsection{Small Structure regime: small groups}

\textsc{hdbscan} detected 14 small groups, defined as those with at least 50 stars independent from any large cluster (i.e. none of the stars are members of any \textit{Big Structure}). Three are located in the surroundings of the $\lambda$ Orionis association, and 11 are in the Orion OB1 association. Most of these do not have a significant star sample with RV information, so a kinematic analysis in the Cartesian phase space cannot be performed.

Four of the 11 clusters in the Orion OB1 association were classified as new clusters: OBP-Centre, OBP-South, OBP-East, and Ori Far West, located in the central region of the OB1 association. Particularly, OBP-East exhibits a wide distribution in the $\mu_{\alpha*}$ component alongside a high dispersion in its CMD that its age estimation, suggesting that OBP-East is composed either by contamination or several minor groups in different evolutionary stages. Additionally, three minor clusters were found in the surroundings of the $\lambda$ Orionis association. Two of these clusters ($\lambda$ Ori South A and $\lambda$ Ori South B) show a proximity of their centroid around $\sim 60$ pc (see Figure~\ref{fig:small_groups_lamori}). When the median of the proper motions from $\lambda$ Orionis is subtracted, they show a likely scenario of future close interactions. However, the lack of RV data makes this statement inconclusive.

\subsection{Small Structure regime: sub-structures}

From our data, five large groups showed signs of sub-structure, defined as small groups arising from the division of \textit{Big Structures}. This classification entails that their stars also belong to their parent main group.

In the OB1 association, we detected six groups as sub-structures in Ori-North, OBP-Near, and Ori-South. The first was divided into the minor clusters $\omega$ Ori and ASCC 20, the second into OBP Near A and OBP Near B and the third into L1647S and L1641. A detailed analysis of these regions in terms of kinematics will be dedicated to future works.

From our analysis, the ONC also showed evidence of sub-structure, revealing three minor clusters: ONC A, ONC B, and ONC C. However, due to the high density of stars, their kinematic features must be analysed carefully. The ONC A group is related to the main structure with more than 600 stars. In comparison, the other two are represented by less than 100 stars with a remarking differentiation in the proper motion orientation but not in parallax.

$\lambda$ Orionis showed three minor groups with \texttt{min\_sample}$=$50. The smallest group corresponded to B30, while the other two belonged to Collinder 69. Another known group, B35, did not appear due to the low amount of stars that it holds. To get the three known clusters (B30, B35 and Collinder 69), we applied \textsc{hdbscan} only to the Group 1 with \texttt{min\_sample}$=$30. This behaviour indicates that, at least in this sample, the low number of objects in B35 is in a regime where \textsc{hdbscan} loses its consistency to preserve larger groups.

\subsection{Close encounters between clusters}

We have identified four potential cluster encounters through kinematic projections. These include two cases in the OB1a association: one involving ASCC 20 and OBP-West in approximately 11.37 Myr, and the other involving $\omega$ Ori and OBP-West in around 1.30 Myr. Additionally, there are two cases in the OBP region: OBP-Far with OBP-d in about 1.03 Myr, and OBP-Far with OBP-b in roughly 2.43 Myr.

\textit{N}-body simulation is needed to understand the dynamical evolution of the clusters located over regions in which cluster--cluster or gas--cluster interactions are undergoing. In future works, a complete assembly of scenarios is planned. The study of the gravitational interaction can also be combined with other astrophysical phenomena to create a model in which young stellar clusters evolve under the conditions found in the Orion Complex.

\section*{Acknowledgements}

We thank the anonymous referee for the comments and suggestions that helped us to improve our paper. S.S.-S. is supported by CONAHCyT Beca Nacional de Posgrado. S.S.-S. and A.P.-V. acknowledge the DGAPA–PAPIIT grant IA103122 and IA103224. C. R-Z, J. H and L.A. acknowledge support from a Group research project, UNAM-DGAPA-PAPIIT IG-101723.
J.B.-P. acknowledges UNAM-DGAPA-PAPIIT support through grant number IN-111-219.
The authors acknowledge CONAHCyT grant 86372 entitled ‘Citlalcóatl: a multiscale study at the new frontier of the formation and early evolution of stars and planetary systems', Mexico.

This work has made use of data from the European Space Agency (ESA) mission
{\it Gaia} (\url{https://www.cosmos.esa.int/gaia}), processed by the {\it Gaia}
Data Processing and Analysis Consortium (DPAC,
\url{https://www.cosmos.esa.int/web/gaia/dpac/consortium}). Funding for the DPAC
has been provided by national institutions, in particular the institutions
participating in the {\it Gaia} Multilateral Agreement.

\section*{Data Availability}

The \textit{Gaia} DR3, APOGEE-2 and GALAH-DR3 data sets used for this work are publicly available. The final data sets for the clusters recovered will be shared upon reasonable request to the corresponding author.



\bibliographystyle{mnras}
\bibliography{main_paper} %




\appendix

\section{Definition and rejection of extended clusters} \label{apendix_b:extended_clusters}

As clusters are expected to form in a compact agglomeration of stars, it would be possible to avoid extended groups by evaluating the distribution of the SIR of the 16th and 86th percentiles from the spatial parameters: $\alpha$, $\delta$ and $\varpi$. It is expected that all real groups are located in a common area on the SIR$_{\alpha}$ versus SIR$_{\delta}$ and SIR$_{\varpi}$ versus SIR$_{\delta}$ projections. Therefore, extended groups should be placed in positions with larger values of SIR$_{\alpha}$, SIR$_{\delta}$, or SIR$_{\varpi}$.

\begin{figure*}
    \centering
    \includegraphics[width=0.4\textwidth]{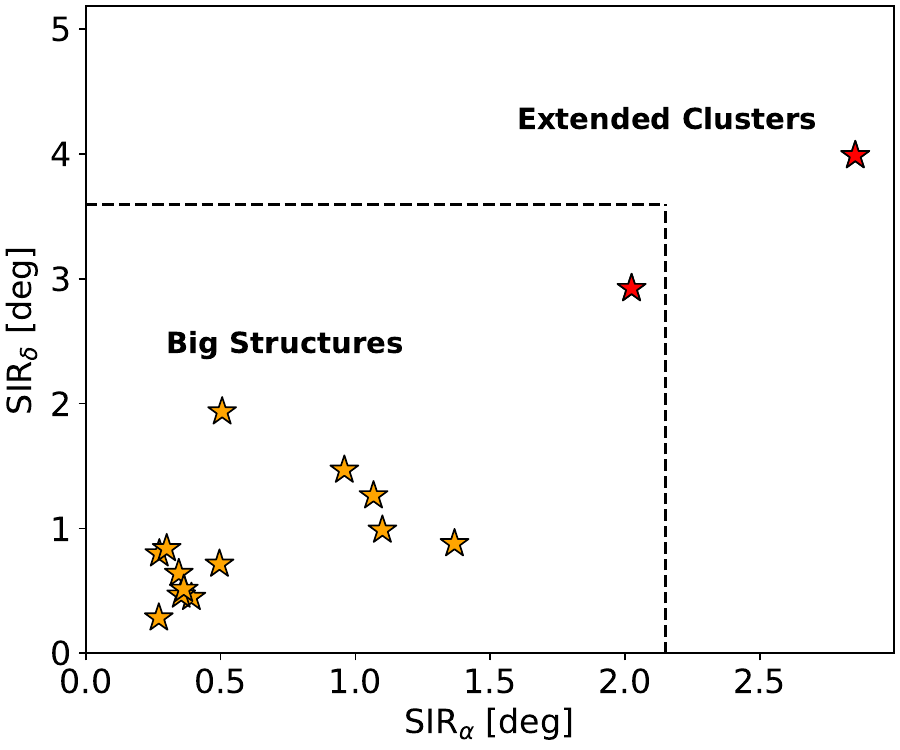}
    \includegraphics[width=0.4\textwidth]{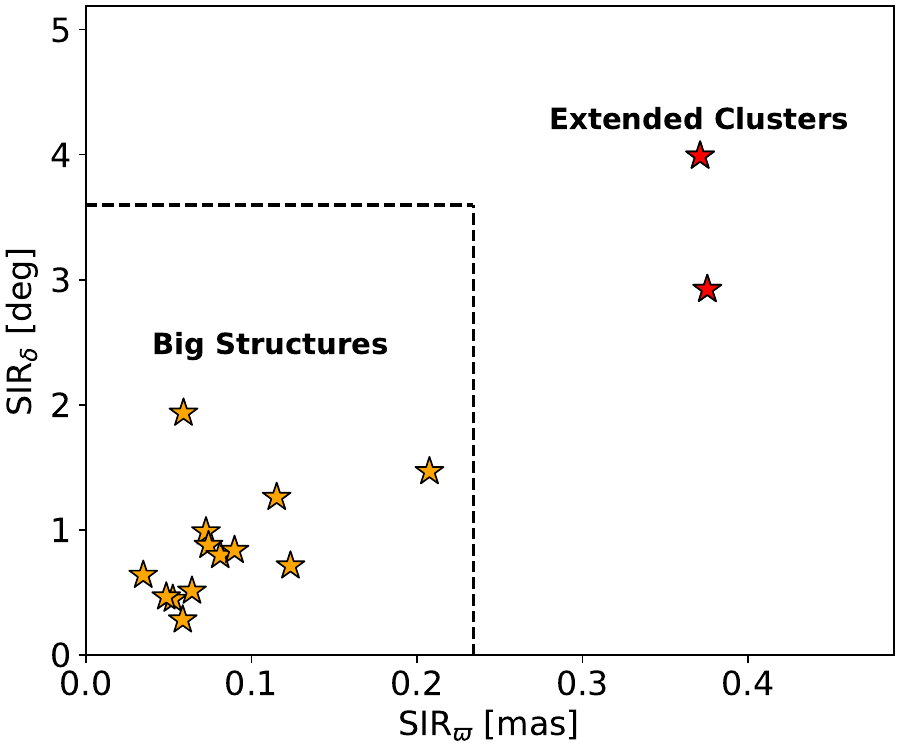}\\
    \includegraphics[width=0.4\textwidth]{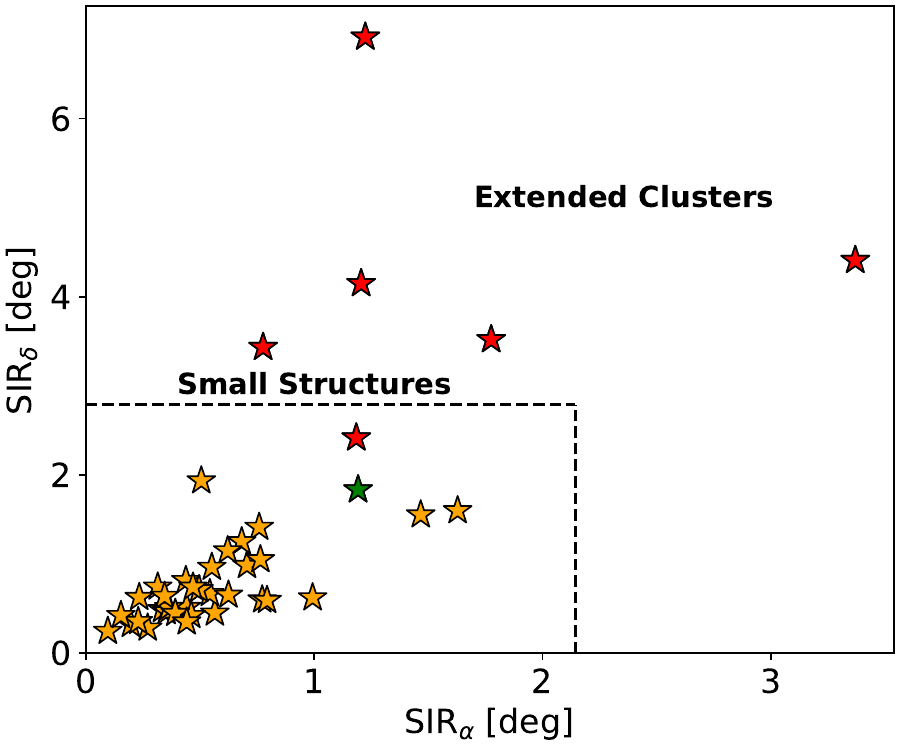}
    \includegraphics[width=0.4\textwidth]{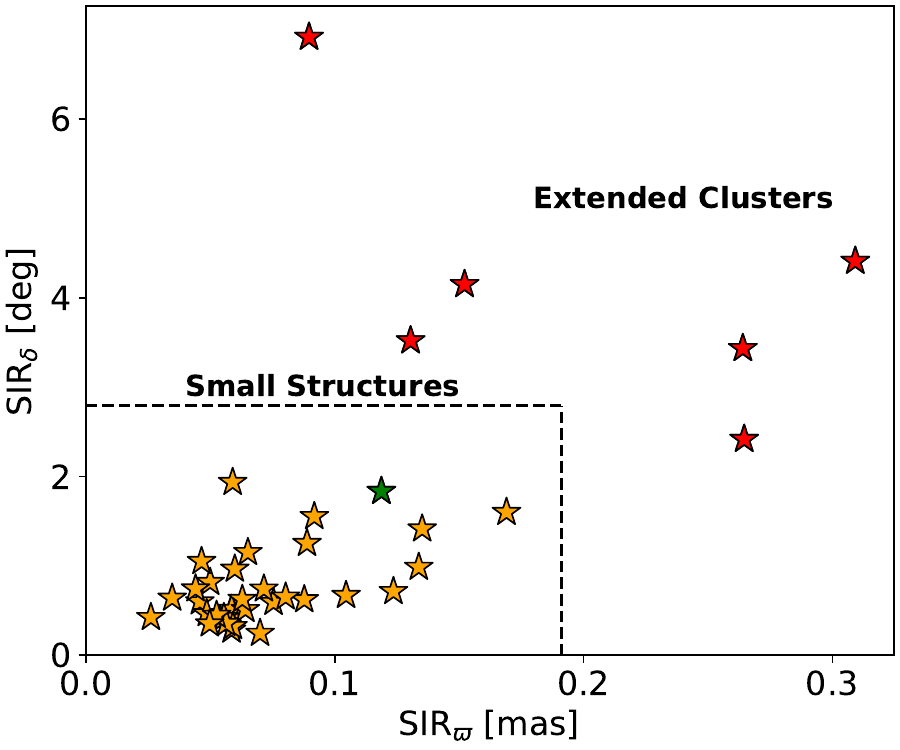}
    \caption{Identification of extended groups according to the distribution of the semi-interquartile range (SIR) in the spatial parameters ($\alpha$, $\delta$, $\varpi$). The dashed lines establishes the $5\sigma$ limit from the median value of every distribution. Red stars correspond to the clusters above the $5\sigma$ limit in any of the projections and the green star in the second row is the special case of OBP-East mentioned in section~\ref{sec:small_groups}. In the first row, we see the result for the \textit{Big Structure} regime in which two groups are rejected. In the second row, we have the analysis for the \textit{Small Structure} regime. This gives a total amount of six groups to be rejected.}
    \label{fig:extended_clusters}
\end{figure*}

In Figure~\ref{fig:extended_clusters}, we show the identification of extended clusters in the \textit{Big Structure} regime and the \textit{Small Structure} regime. In all cases, we identify as extended groups those that appear outside the limit of $5\sigma$ from the median distribution of points using MAD as a statistical parameter for the dispersion. From the \textit{Big Structure} regime, we got one extended cluster in the SIR$_{\alpha}$ versus SIR$_{\delta}$ projection and an additional cluster in the SIR$_{\varpi}$ versus SIR$_{\delta}$ projection, with both rejected from the results. On the other hand, in the \textit{Small Structure} regime, five clusters were considered as extended in the SIR$_{\rm{\alpha}}$ versus SIR$_{\rm{\delta}}$ projection and one additional cluster in the SIR$_{\rm{\varpi}}$ versus SIR$_{\rm{\delta}}$ projection.

\section{Extended Groups}
\label{apendix_c:extended}

\begin{figure*}
    \centering
    \includegraphics[width=0.95\textwidth]{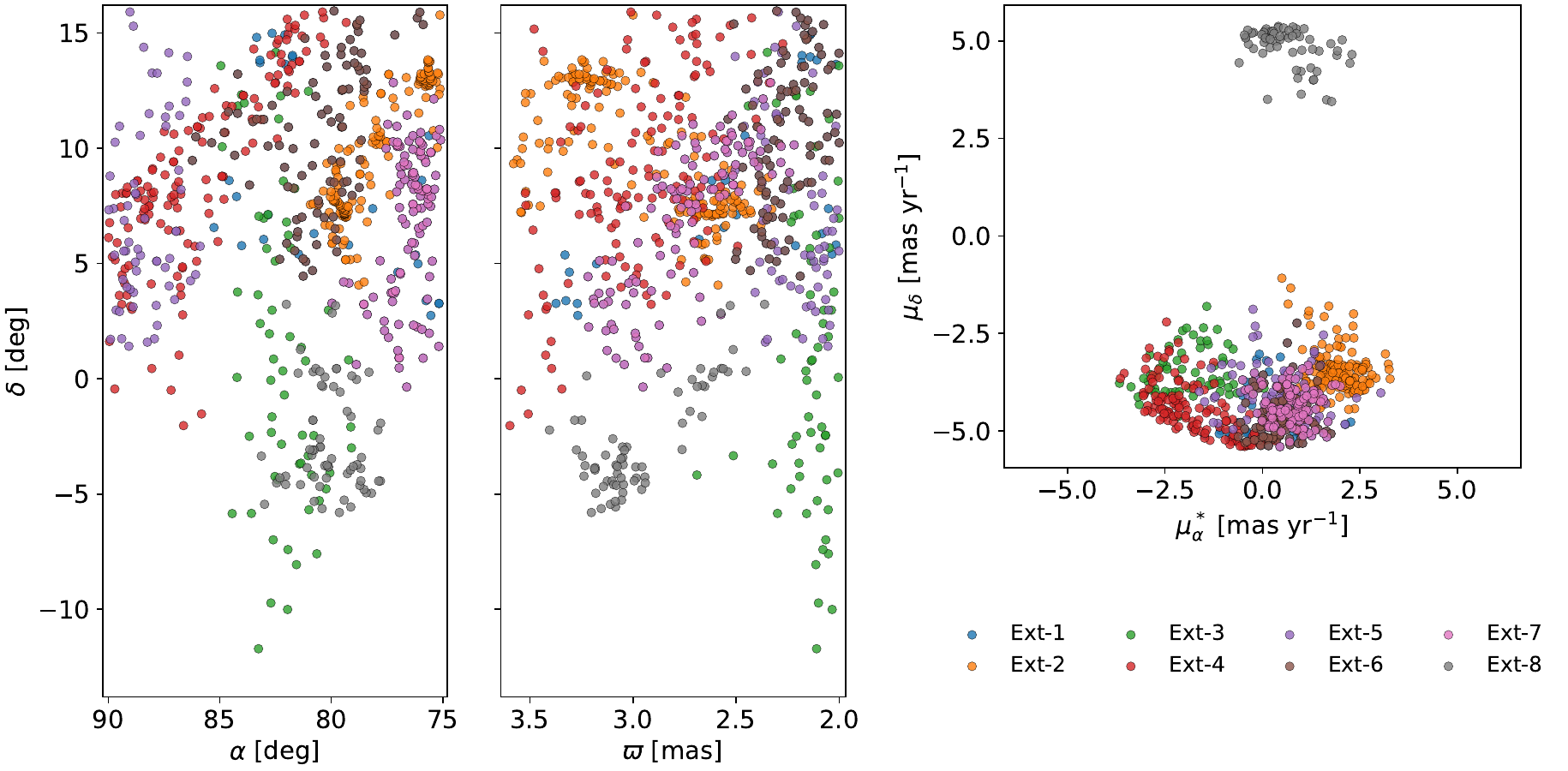}
    \caption{Distribution of the recovered extended groups in the Orion Complex. \textit{Left panel}: distribution of the groups in the sky projection. \textit{Middle panel}: distribution in $\delta$ versus $\varpi$. \textit{Right panel}: vector--point diagram.}
    \label{fig:extended_clusters_2}
\end{figure*}

In Table~\ref{table:extended}, we provide the astrometric information of the eight extended groups detected by \textsc{hdbscan} but rejected for the analysis. Additionally, in Figure~\ref{fig:extended_clusters_2}, we show the distribution of these clusters in the sky projection (left panel), $\varpi$ versus $\delta$ (middle panel) and the vector--point diagram (right panel). Note that most of the stars are mainly spread in the northern region of the OSFC ($\delta > 0^{\circ}$). Similarly, the majority of the extended groups are notably located in the vector--point diagram over ranges $-4.0<\mu^*_{\alpha}<4.0$ and $-6.0<\mu_{\delta}<-1.0$ mas yr$^{-1}$.

{\renewcommand{\arraystretch}{1.24}
\begin{table*}
\centering
\caption{Astrometric information from the extended groups provided by \textit{Gaia}. The median of every quantity is given alongside the 16th and 84th percentiles. In the second column, the regime is specified with BS for \textit{Big Structure} and SS for \textit{Small Structure}.}
\label{table:extended}
\begin{tabular}{cccccccc}
\hline
Cluster & \multicolumn{1}{c}{Regime} & \begin{tabular}[c]{@{}c@{}}$\bar{\alpha}$\\ {(}deg{)}\end{tabular} & \begin{tabular}[c]{@{}c@{}}$\bar{\delta}$\\ {(}deg{)}\end{tabular} & \begin{tabular}[c]{@{}c@{}}$\bar{\varpi}$\\ {(}mas{)}\end{tabular} & \begin{tabular}[c]{@{}c@{}}$\bar{\mu}_\alpha^*$\\ {(}$\rm{mas}\ s^{-1}${)}\end{tabular} & \begin{tabular}[c]{@{}c@{}}$\bar{\mu}_\delta$\\ {(}$\rm{mas}\ s^{-1}${)}\end{tabular} & \begin{tabular}[c]{@{}c@{}}$N_T$\\ \end{tabular} \\ \hline\hline
Ext-1  & BS  & 77.72$^{+3.94}_{-1.77}$ & 8.56$^{+4.00}_{-3.98}$ & 2.46$^{+0.45}_{-0.29}$ & 0.58$^{+0.55}_{-0.61}$ & -4.53$^{+0.67}_{-0.52}$ & 230    \\
Ext-2  & BS  & 78.61$^{+1.09}_{-2.96}$ & 9.34$^{+3.65}_{-2.19}$ & 2.95$^{+0.37}_{-0.38}$ & 1.96$^{+0.46}_{-0.47}$ & -3.53$^{+0.43}_{-0.35}$ & 157  \\
Ext-3  & SS  & 82.27$^{+1.13}_{-1.32}$ & 0.34$^{+8.04}_{-5.79}$ & 2.13$^{+0.10}_{-0.08}$ & -1.89$^{+0.94}_{-0.85}$  & -3.67$^{+0.89}_{-0.34}$ & 61 \\
Ext-4  & SS  & 86.67$^{+2.31}_{-4.43}$ & 8.57$^{+5.12}_{-3.69}$ & 2.97$^{+0.34}_{-0.27}$ & -1.96$^{+1.43}_{-0.69}$  & -4.50$^{+0.74}_{-0.70}$ & 134 \\
Ext-5  & SS  & 88.55$^{+0.81}_{-1.60}$ & 5.92$^{+5.71}_{-2.59}$ & 2.19$^{+0.18}_{-0.13}$ & 0.29$^{+1.04}_{-1.05}$ & -4.11$^{+0.83}_{-0.61}$ & 61   \\
Ext-6  & SS  & 88.17$^{+2.14}_{-1.41}$ & 10.80$^{+3.24}_{-3.80}$ & 2.27$^{+0.12}_{-0.14}$  & 0.39$^{+0.54}_{-0.42}$ & -4.70$^{+0.77}_{-0.49}$ & 92   \\
Ext-7  & SS  & 76.34$^{+0.89}_{-0.66}$  & 7.93$^{+2.17}_{-4.70}$ & 2.73$^{+0.29}_{-0.23}$  & 0.76$^{+0.50}_{-0.50}$ & -4.40$^{+0.54}_{-0.42}$ & 111  \\
Ext-8  & SS  & 80.19$^{+0.87}_{-1.50}$  & -3.41$^{+3.63}_{-1.20}$ & 3.05$^{+0.13}_{-0.39}$  & 0.45$^{+0.86}_{-0.43}$ & 5.04$^{+0.20}_{-0.71}$ & 65   \\ \hline
\multicolumn{8}{l}{}\\
\end{tabular}
\end{table*}
}

\bsp	
\label{lastpage}
\end{document}